\documentclass[
aps,
prx,
groupedaddress,
superscriptaddress,
floatfix,
twocolumn
]{revtex4-1}

\pdfoutput=1

\usepackage{graphicx}
\usepackage{amsmath,amssymb}
\usepackage{hyperref} 
\usepackage{bm} 
\usepackage{mathtools} 
\usepackage{float}
\usepackage{braket}

\newif\iffigures
\figurestrue

\newcommand{\p}{\partial}
\newcommand{\ep}{\varepsilon}

\newcommand{\ka}{\kappa}
\newcommand{\al}{\alpha}
\newcommand{\la}{\lambda}

\renewcommand\Re{\operatorname{Re}}
\renewcommand\Im{\operatorname{Im}}

\begin{document}

\title{Topological spin Meissner effect in exciton-polariton spinor condensate: constant amplitude solutions, half-vortices and symmetry breaking}

\author{D. R. Gulevich} \email{drgulevich@metalab.ifmo.ru}
\affiliation{ITMO University, St. Petersburg 197101, Russia}
\affiliation{Department of Physics, University of Bath, Bath BA2 7AY, United Kingdom}

\author{D. V. Skryabin} 
\affiliation{ITMO University, St. Petersburg 197101, Russia}
\affiliation{Department of Physics, University of Bath, Bath BA2 7AY, United Kingdom}

\author{A. P. Alodjants}
\affiliation{ITMO University, St. Petersburg 197101, Russia}

\author{I. A. Shelykh}
\affiliation{ITMO University, St. Petersburg 197101, Russia}
\affiliation{Science Institute, University of Iceland, Dunhagi 3, IS-107, Reykjavik, Iceland}
\affiliation{Division of Physics and Applied Physics, Nanyang Technological University 637371, Singapore}

\date{\today}

\begin{abstract} 
We generalize the spin Meissner effect for exciton-polariton condensate confined in annular geometries to the case of non-trivial topology of the condensate wavefunction.
In contrast to the conventional spin Meissner state, topological spin Meissner states can in principle be observed at arbitrary high magnetic field not limited by the critical magnetic field value for the condensate in a simply-connected geometry. One special example of the topological Meissner states are half-vortices. 
We show that in the absence of magnetic field half-vortices in a ring exist in a form of superposition of elementary half-vortex states which resolves recent  experimental results where such puzzling superposition was observed.
Furthermore, we show that if a pure half-vortex state is to be observed, a non-zero magnetic field of a specific magnitude needs to be applied.
Studying exciton-polariton in a ring in presence of TE-TM splitting, we observe spin Meissner states which break rotational symmetry of the system by developing inhomogeneous density distributions. We classify various states arising in presence of non-zero TE-TM splitting based on what states they can be continued from by increasing the TE-TM splitting parameter from zero. With further increasing TE-TM splitting, states with broken symmetry may transform into stable half-dark solitons and therefore may serve as a useful tool to generate various non-trivial states of a spinor condensate.
\end{abstract}

\maketitle
\tableofcontents

\section{Introduction}
\label{section:Intro}

Development of nanotechnology achieved during the last decade allowed the design of semiconductor microcavities possessing extremely high Q--factors (more than 10000). This opened new opportunities in creation and investigation of fundamental properties of Bose-Einstein condensates (BEC) of exciton-polaritons -- hybrid light-matter quasiparticles emerging in the regime of strong coupling~\cite{Hui}. Although real thermodynamic equilibrium in polariton condensates is never achieved and thus they are fundamentally different from atomic BECs, they exhibit main properties inherent to weakly interacting quantum Bose-gases.  Among them is superfluidity ~\cite{Carusotto,AmoSuperfluidity}, formation of quantized vortices ~\cite{Lagoudakis} and solitons \cite{Sich}, Josephson oscillations and macroscopic self trapping \cite{AmoSelfTrapping}, spin-Hall effect ~\cite{OSHENPhys} and others. The peculiarity of spin structure of polaritons combined with strong polariton-polariton interactions and large coherence lengths makes possible the generation of coherent bosonic spin currents ~\cite{Shelykh-SST-2010}. This opens a new research field of light-mediated spin effects and paves the way for their implementation in optoelectronics, e.g. for the creation of all-optical integrated circuits ~\cite{Ballarini,Sturm-Nature-2014}.

Polariton systems possess sevaral advantages with respect to systems based on cold atoms. First, extremely small mass of the polaritons (about $10^{-5}$ of the mass of free electrons) makes critical temperatures of the observation of quantum collective effects surprisingly high (from a few Kelvin for GaAs based structures to room temperature for GaN structures \cite{RoomTemp}). Besides, polariton condensates allow reasonably simple manipulation by application of the external electric and magnetic fields ~\cite{Schneider, SchneiderSciRep}. This plays essential role in study of the fundamental properties of exciton-polariton condensates. In general case magnetic field affects exciton-polariton emission energy, linewidth and intensity due to the exciton energy shift caused by Zeeman splitting in circular polarizations, modification of the exciton-photon coupling strength \cite{ShelykhIvchenko}, and modification of  scattering process with acoustic phonons ~\cite{Pietka}.  Strong spin anisotropy of polariton-polariton interactions, however, make those dependencies in the non-linear regime highly non-trivial.  In particular,  in Ref.~\cite{Rubo-PLA-2006} it was shown that below some critical value $B_{c}^{} $ of magnetic field  depending on polariton concentration the so-called full paramagnetic screening (also known as spin Meissner effect) occurs. Its signature is independence of the photoluminescence energy on the magnetic field. The latter however affects the polarization of the emission. Its ellipticity gradually changes until the value $B_{c}^{}$ is reached. At this point emission becomes fully circular polarized and Zeeman splitting re-establishes. Main efforts of recent experimental studies of exciton-polariton condensates in magnetic field have been successfully directed towards confirmation of these seminal peculiarities, cf. ~\cite{Pietka,Larionov-PRL-2010,Sturm-PRB-2015,Walker-PRL-2011,Fisher-PRL-2014}.

The interplay between polarization splitting and anisotropic polariton-polariton interactions becomes more tricky in anisotropic cavities when  additional energy splittings in linear polarizations (TE-TM splittings) is present in addition to the Zeeman splitting~\cite{Shelykh-Superlattices}. The situation becomes even more interesting when polaritons are confined in non-simply connected region, e.g. inside ring resonator. In this case the direction of the effective magnetic field provided by TE-TM splitting becomes position-dependent which combined together with magnetic field induced Zeeman splitting leads to the appearance of the geometric Berry phase responsible for generation of synthetic U(1) gauge field for polaritons and possibility of observation of optical analog of Aharonov-Bohm effect ~\cite{Shelykh-PRL-2009}. It should be noted that exciton-polariton spinor BEC in a ring geometry have been experimentally demonstrated by several groups ~\cite{Sturm-Nature-2014, Larionov-PRL-2010, Sturm-PRB-2015, Walker-PRL-2011, Liu-PNAS-2015}. However, polarization properties of interacting spinor polaritons in the rings were not subject of theoretical investigation up to now for the best of our knowledge. On the other hand, the presence of artificial U(1) gauge potential can lead to the onset of the persistent current in the system, i.e. its ground state can be a vortex-type solution. The investigation of the analogs of spin Meissner effect for such states with quantized angular momentum is a fundamentally interesting problem which can in principle lead to applications such as polariton analogue of flux qubits.

Spinor vortex-type solutions in 2D systems were analyzed in Ref.~\cite{Rubo-PRL-2007}. It was shown that besides normal vortices for which both circular polarization components have same non-zero quantized angular momentum, half-vortex solutions for which one of the circular polarizations is not rotating. Half-vortices had been detected experimentally~\cite{Lagoudakis-Science-2009}, but their stability in 2D condensate in presence of TE-TM splitting remained a topic of a debate~\cite{Flayac-PRB-2010, Solano-Rubo-comment, Flayac-reply}. The current view is that small TE-TM splitting does not destroy half-vortices leading only to their warping ~\cite{Solano-PRB-2014}. However, large TE- TM splittings can make in principle half vortex solutions instable. This situation may become relevant when polaritons are confined in the ring, where relevant splittings can reach the values of 1-2 meV for ring thicknesses about 1 micron \cite{Kuther-PRB-1998}.  Recent experimental work on half-vortices in a ring geometry~\cite{Liu-PNAS-2015} demonstrated that polarization patterns and density profiles could not be explained by the existing theory, which led the authors to a conclusion that their experimental configuration corresponds to some spurious superposition of certain "elementary" half-vortex states.

The aim of this paper is to provide a complete theory of interacting non-simply connected polariton BEC, introduce a concept of topological spin Meissner effect and describe rich variety of the states of the condensate both in presence and absence of the TE-TM splitting. In particular, we provide detailed analysis of half-vortex states in the ring and show that superpositions of elementary half-vortex states reported experimentally in Ref.~\cite{Liu-PNAS-2015} appear naturally in the developed theory.

The paper is organized as follows. In Section II we present the model of exciton-polariton condensate in a ring and introduce various types of emerging solutions. In Section III we introduce the topological spin Meissner effect in the case when TE-TM splitting is absent. In Section IV and V we extend the concepts and solutions obtained in the previous section to the general case of finite TE-TM splitting. More specifically, Section IV deals with topological spin Meissner states in the form of constant amplitude solutions and in Section V we study states which spontaneously break the rotational symmetry of the system due to presence of TE-TM splitting. Section VI contains discussion of the experimental relevance of our parameters. 

\section{Model and classification of solutions}
\label{sec:model}

Interacting polaritons trapped in a quasi one-dimensional ring resonator  can be described by the following system of dimensionless Gross-Pitaevskii equations (see Appendix~\ref{app:tetm}),
\begin{equation}
\begin{dcases}
i\dot\psi_+= - \partial_x^2\psi_+ + \left( |\psi_+|^2 + \alpha |\psi_-|^2 \right)\psi_+ \\
\quad\quad\quad\quad\quad\quad\quad + \Omega\psi_+  + \kappa e^{-2 i x} \psi_-,
\\
i\dot\psi_- = - \partial_x^2\psi_- + \left( |\psi_-|^2 + \alpha |\psi_+|^2  \right)\psi_- 
\\
\quad\quad\quad\quad\quad\quad\quad - \Omega\psi_- + \kappa e^{2 i x} \psi_+.
\end{dcases}
\label{GP-ring}
\end{equation}
Here, $\psi_{\pm}$ are the components of the  exciton-polariton spinor wavefunction $\pmb\psi\equiv\{\psi_+,\psi_-\}$ in the basis of circular polarizations satisfying $\psi_{\pm}(t,x)=\psi_{\pm}(t,x+2\pi)$, parameter $\alpha<0$ characterizes attractive interaction of the cross-polarized polaritons, $\Omega$ is half of Zeeman splitting of a free polariton state (which we will refer to as just "magnetic field") and $\kappa$ is half of the momentum independent TE-TM energy splitting. Parameters $\Omega$ and $\kappa$ are dimensionless and scale in units of $\hbar^2/(2 m^* R^2)$, where $R$ is the ring radius  and $m^*$ is the exciton-polariton effective mass.
We use the dimensionless  particles density per unit length, $\rho\equiv \frac{1}{2\pi}\int_0^{2\pi}\left(|\psi_+|^2+|\psi_-|^2\right)dx$, as a parameter controlling strength of the polariton-polariton interactions. 

To study stationary states of the system~\eqref{GP-ring} we use the substitution 
\begin{equation}
\psi_{\pm}(t,x) = 
\psi_{\pm}(x) e^{-i\mu t}.
\end{equation}
We treat $\mu$ as an unknown variable corresponding to the energy blue shift of a photoluminescence line of the condensate in a steady state~\cite{Kavokin-Microcavities},  found for a given $\rho$. It is also identical to the chemical potential
parameter used in the literature on the Gross-Pitaevskii model and Bose-Einstein condensation. When time-dependent problems are treated $\mu$ gives a frequency of the rotating frame in which dynamics of physical quantities is captured. 

The system of equations~\eqref{GP-ring} inherits properties of the nonlinear Schr\"{o}dinger equation. Similar to the nonlinear Schr\"{o}dinger equation~\cite{Carr-I}, an important class of stationary solutions of Eqs.~\eqref{GP-ring} are constant amplitude solutions, which correspond to polarization vortices with the homogeneous density distributions along the ring. Those can be sought in the form 
\begin{equation}
\psi_{\pm}(x)=\chi_{\pm}e^{im_{\pm}x},\quad\text{for}\quad~\kappa =0.
\label{const-amp-sols}
\end{equation}
and
\begin{equation}
\psi_{\pm}(x)=\chi_{\pm}e^{i(n\mp 1)x},\quad\text{for}\quad~\kappa\ne 0 
\label{const1}
\end{equation}
Here, $\chi_{\pm}$ are $x$-independent amplitudes. 
Therefore in the presence of the TE-TM splitting vortex winding numbers in two components of the spinor must 
differ by $2$, while these winding numbers $m_{\pm}$ are arbitrary integers for ${\kappa=0}$.
In the  limit of noninteracting polaritons $\mu$ gives the energy spectrum and the existence of solutions \eqref{const1} and \eqref{const-amp-sols} requires 
\begin{equation}
\mu =\mu_\pm^{(0)},\; ~\mu_\pm^{(0)}= m_{\pm}^2 \pm \Omega
\label{linear-zero}
\end{equation}
and
\begin{eqnarray}
 && \mu =\mu_\pm,\; ~\mu_\pm = 1+n^2 \pm \sqrt{(2n-\Omega)^2+\kappa^2},
\label{linear-kappa}
\end{eqnarray}
respectively.
It is clear that the vortex energies in the no-interaction limit vary  linearly  with the applied magnetic field $\Omega$ for $\kappa= 0$, While for $\kappa\ne 0$, one deals with the typical anticrossing behavior in the proximity of the points $\Omega=2n$ (see detailed discussion and figure in Sec.~IV). 

When nonlinear effects are included, the  energies acquire the corresponding nonlinear shifts proportional to~$\rho$, but not only this. Spin anisotropy of nonlinear interaction $\alpha\ne 1$ makes it possible for the mixed ($\chi_{+}\ne 0, \chi_-\ne 0$) vortex states to loose the dependence of their energies on the applied magnetic field. Such behavior of exciton-polariton condensate in thermodynamic equilibrium is known as spin Meissner effect (for introduction to the spin Meissner effect see Section~\ref{section:Intro} and original Refs.~\cite{Rubo-PLA-2006, Shelykh-Superlattices, Larionov-PRL-2010}). 
However, in contrast to the previously studied spin Meissner effect, properties of such vortex states and their domain of existence are defined by the vortex winding numbers. 
To highlight the dependence on the winding numbers  we term the vortex states whose energies either exactly or approximately lose dependence on the magnetic field -- {\it topological spin Meissner states} (TSM states). As it will be shown below TSM states are ubiquitous feature of our model.
Note, that vortices  with  $m_+=\pm 1$, $m_-=0$ and $m_+=0$, $m_-=\pm 1$ are so-called half-vortices in the terminology used in \cite{Rubo-PRL-2007,Lagoudakis-Science-2009,Flayac-PRB-2010,Solano-Rubo-comment,Flayac-reply,Solano-PRB-2014}. We will show that half-vortices also exhibit topological spin Meissner effect.

We study nonlinear solutions for $\kappa=0$ in Section III. Importantly, solutions~\eqref{const-amp-sols} with $m_--m_+\ne 2$,  do not disappear as we introduce $\kappa\ne 0$, they simply develop  inhomogeneous density profiles and thus are associated with the breaking the rotational symmetry. We study the $\kappa\ne 0$ case in details in  Section IV. Note, that the system of Eq.~\eqref{GP-ring} even with $\kappa=0$  has various soliton-like solutions with the inhomogeneous density profiles, see, e.g.~\cite{Carr-I}. These solitons can continue to exist for non-zero $\kappa$ as well. In order to limit the scope of the present work we leave these type of inhomogeneous solutions for future studies.

\section{Topological spin Meissner effect: Zero TE-TM splitting.}
\label{section:zero}

\subsection{Stationary Solutions}
We first focus on the case when the TE-TM splitting is absent, $\kappa=0$.
Substituting~ Eq.~\eqref{const-amp-sols} to~\eqref{GP-ring}, 
we have,
\begin{eqnarray}
\nonumber && \left[-\mu+m_+^2 + \chi_{+}^2+\alpha\chi_{-}^2 +\Omega\right]\chi_{+}=0,\\
&& \left[-\mu+m_-^2 + \chi_{-}^2+\alpha\chi_{+}^2 -\Omega\right]\chi_{-}=0.
\label{chi-eq-general}
\end{eqnarray}
Because the phases of $\psi_\pm$ are arbitrary, in this section we will assume $\chi_-,\chi_+\ge 0$ without loosing the generality, however the relative phase of the amplitudes will play an important role when we will be dealing with the case of non-zero TE-TM splitting in the next section.
One obvious class of solutions of Eqs.~\eqref{chi-eq-general} comes by setting either $\chi_{+}$ or $\chi_{-}$ to zero: this gives two solutions with amplitudes $(\chi_{+},\chi_{-})=(\sqrt{\rho},0)$ and $(\chi_{+},\chi_{-})=(0,\sqrt{\rho})$ 
and chemical potentials,
\begin{equation}
\mu_+^{(0)}=m_+^2+\rho+\Omega
\label{mu-plus}
\end{equation}
and
\begin{equation}
\mu_-^{(0)}=m_-^2+\rho-\Omega,
\label{mu-minus}
\end{equation}
respectively. Energies of these solutions  either increase or decrease with $\Omega$, depending on whether the polariton spin is parallel or anti-parallel to the applied magnetic field. 

\iffigures
\setlength{\unitlength}{0.1in}
\begin{figure}
\begin{center}
$
\begin{array}{c}
\begin{picture}(35,23)
\put(-1,0){\includegraphics[width=3.4in]{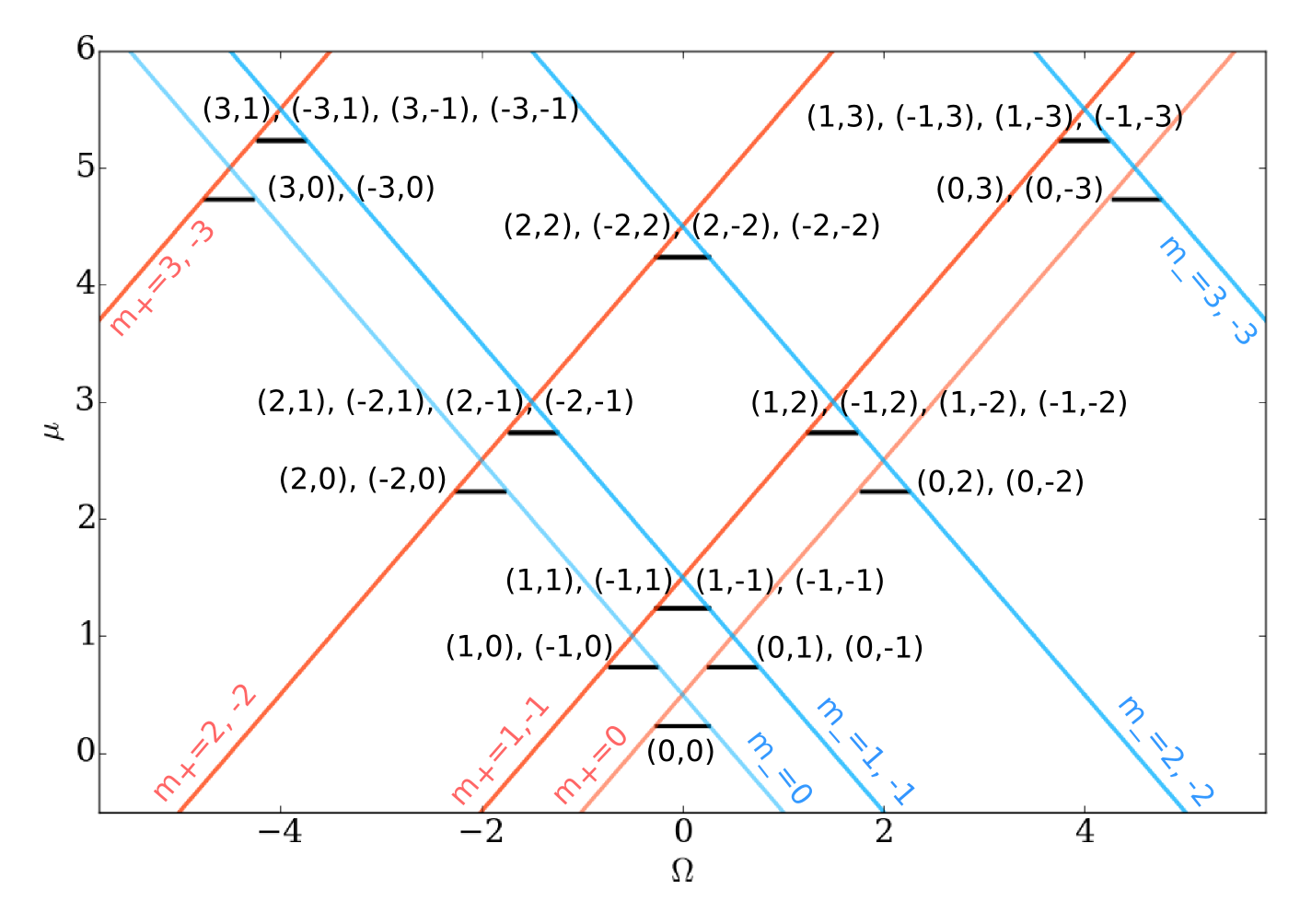}}
\put(2,20.5){(a)}
\end{picture}
\\
\begin{picture}(35,23)
\put(-1,0){\includegraphics[width=3.4in]{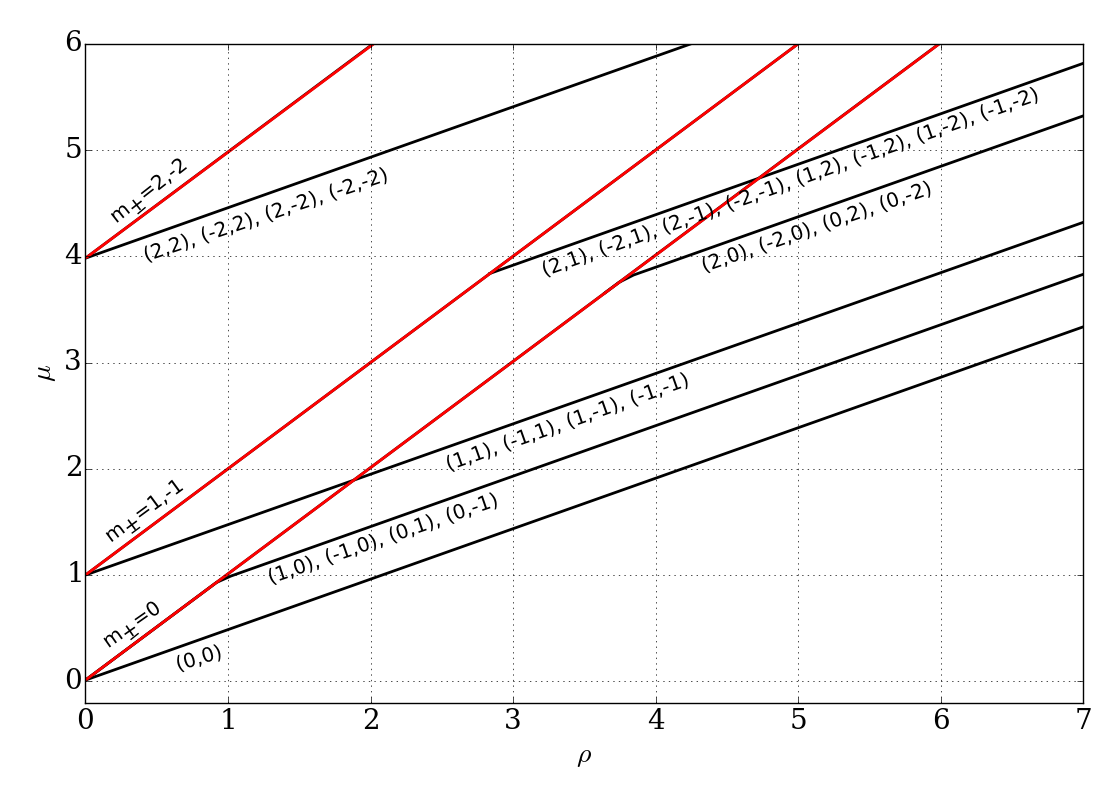}}
\put(2,21){(b)}
\end{picture}
\end{array}
$
\caption{\label{fig:families} 
(Color online) Families of constant amplitude solutions~\eqref{const-amp-sols} in zero TE-TM splitting ($\kappa=0$) on the diagrams  $\mu$ vs $\Omega$ at fixed $\rho=0.5$ (a) and $\mu$ vs $\rho$ at fixed $\Omega=0$. Red and blue solid lines are pure circular polarization vortices characterized by winding numbers $m_\pm$ and chemical potential given by Eqs.~\eqref{mu-plus} and~\eqref{mu-minus}. Black lines are topological spin Meissner (TSM) states specified by a pair of winding numbers $(m_+,m_-)$. Each TSM state appear from interaction of circular polarization vortices and if nonlinearity is increased from zero arise from the corresponding intersections of the red and blue lines on Figure (a). Figure (b) shows how TSM states appear on the diagram $\mu$ vs $\rho$ when nonlinearity parameter is increased continuously from $\rho=0$ (linear case) to $7$ in zero magnetic field $\Omega=0$. 
In both cases $\alpha=-0.05$.
}
\end{center}
\end{figure}
\fi

\iffigures
\setlength{\unitlength}{0.1in}
\begin{figure}
\begin{center}
\includegraphics[width=3.4in]{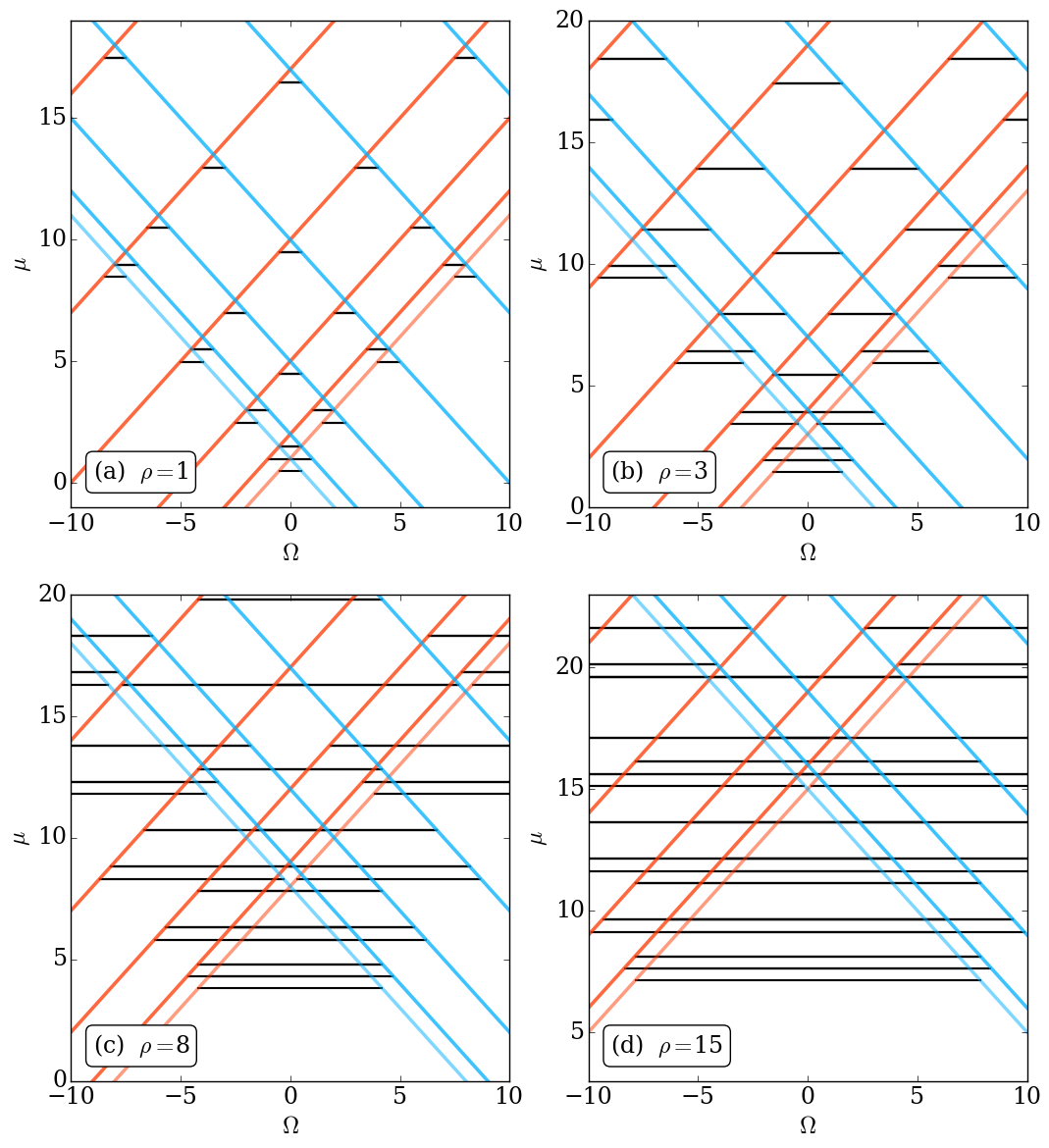}
\caption{\label{fig:largescale} 
(Color online) 
Constant amplitude solutions~\eqref{const-amp-sols} in zero TE-TM splitting ($\kappa=0$) on the diagrams  $\mu$ vs $\Omega$ at different values of the nonlinearity parameter $\rho=1$ (a), $3$ (b) $8$ (c) and $15$ (d). 
Topological spin Meissner (TSM) states are marked by black lines resting on the corresponding red and blue diagonal lines corresponding to the pure circular polarization vortices. The intervals of magnetic field where a given TSM state exists grow with increasing the nonlinearity $\rho$. More details are provided in Fig.~\ref{fig:families} where the specific TSM states are marked by their winding numbers. In all cases $\alpha=-0.05$.
}
\end{center}
\end{figure}
\fi

\iffigures
\setlength{\unitlength}{0.1in}
\begin{figure}
\begin{center}
$
\begin{array}{cc}
\begin{picture}(16,18)
\put(-2,0){\includegraphics[width=1.75in]{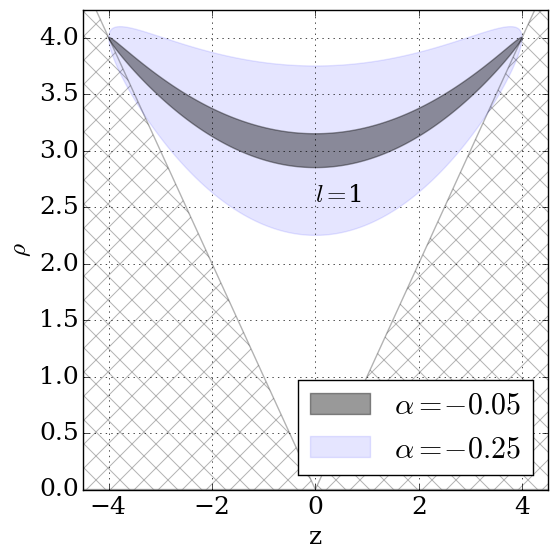}}
\put(1,3){(a)}
\end{picture}
&
\begin{picture}(16,18)
\put(-1,0){\includegraphics[width=1.75in]{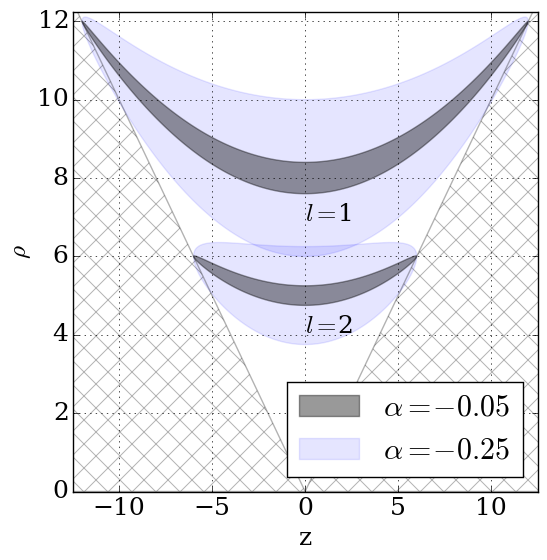}}
\put(2,3){(b)}
\end{picture}
\\
\begin{picture}(16,18)
\put(-2,0){\includegraphics[width=1.75in]{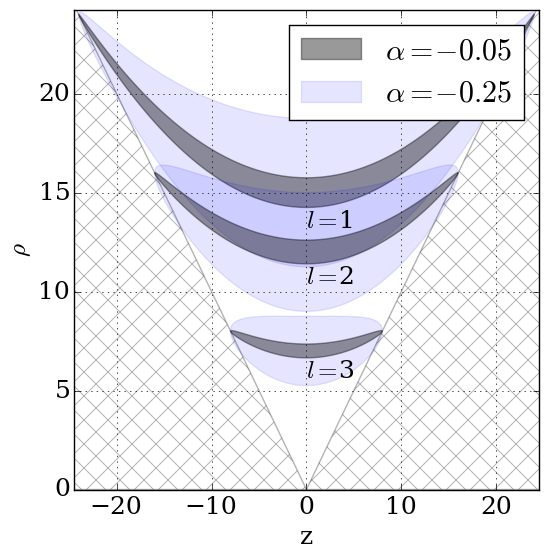}}
\put(1,3){(c)}
\end{picture}
&
\begin{picture}(16,18)
\put(-1,0){\includegraphics[width=1.75in]{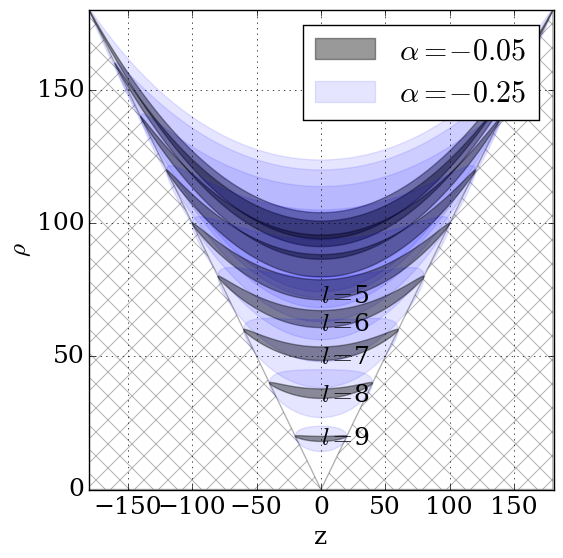}}
\put(2,3){(d)}
\end{picture}
\end{array}
$
\caption{\label{fig:z-rho} 
(Color online)
Instability regions of topological spin Meissner (TSM) states with $|\Delta m|=2$ (a), $|\Delta m|=3$ (b), $|\Delta m|=4$ (c) and $|\Delta m|=10$ (d) as given by the analytical formulas~\eqref{rhoc} and~\eqref{Delta-rho}. Dark and light blue areas are instability regions in parameter space $(z,\rho)$ arising at different angular harmonics $|l|$ for two different values of the parameter $\alpha$: $-0.05$ and $-0.2$, correspondingly. The hatched area marks the region $|z|>\rho$ where solutions in the form of TSM states do not exist (see~\eqref{condition-Meissner}).
}
\end{center}
\end{figure}
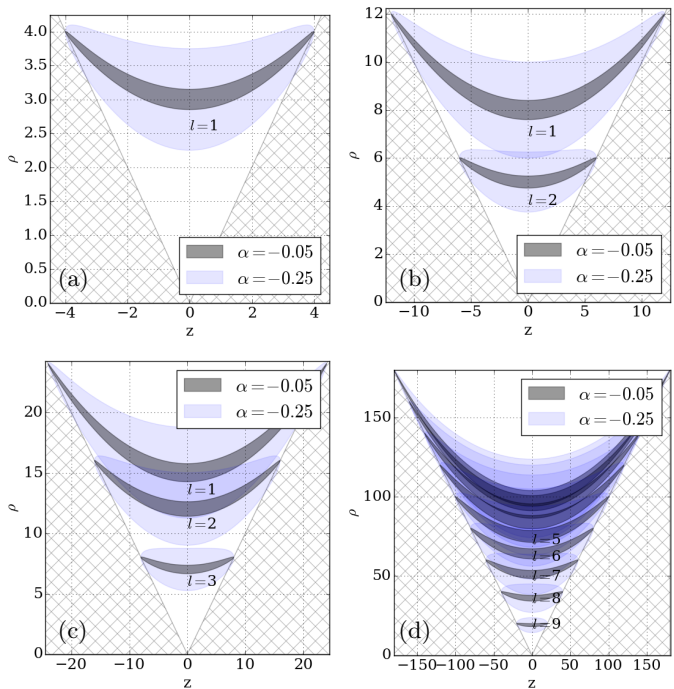
\fi

\iffigures
\setlength{\unitlength}{0.1in}
\begin{figure*}
\begin{center}
$
\begin{array}{cc}
\begin{picture}(35,22)
\put(0,0){\includegraphics[width=3.5in]{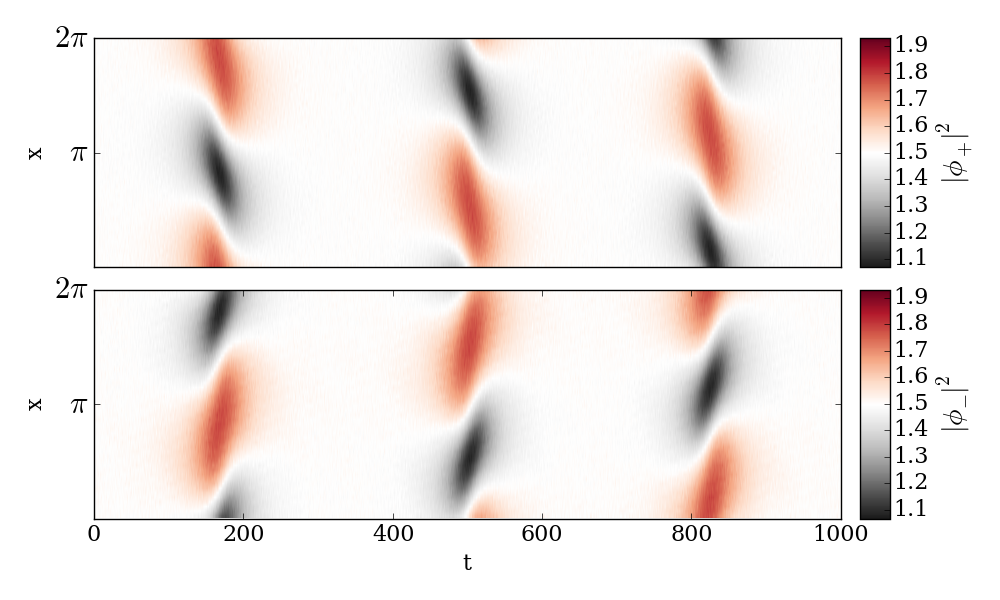}}
\put(3,0){(a)}
\end{picture}
&
\begin{picture}(35,22)
\put(0,0){\includegraphics[width=3.5in]{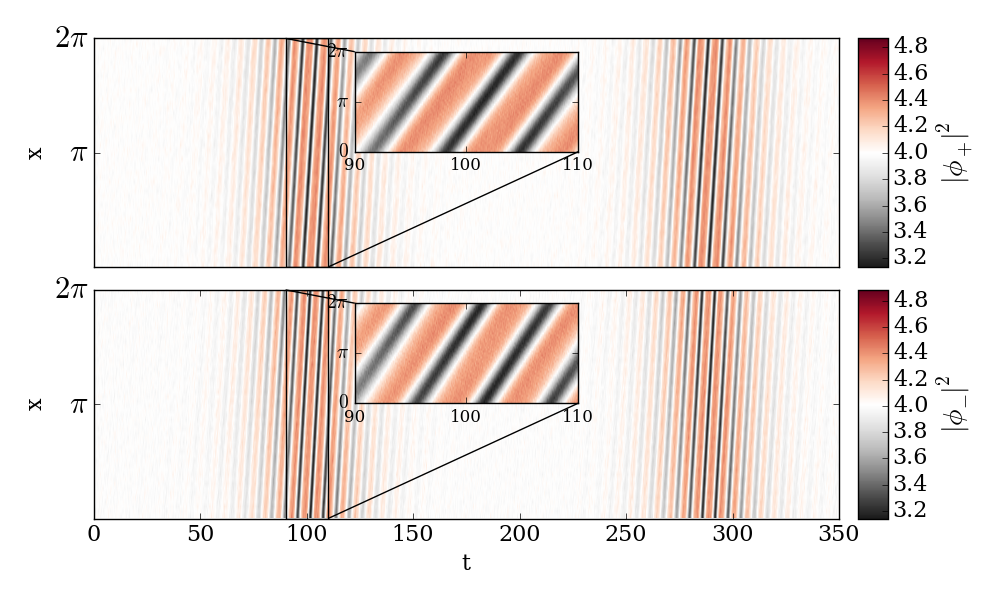}}
\put(3,0){(c)}
\end{picture}
\\
\begin{picture}(35,22)
\put(0,0){\includegraphics[width=3.5in]{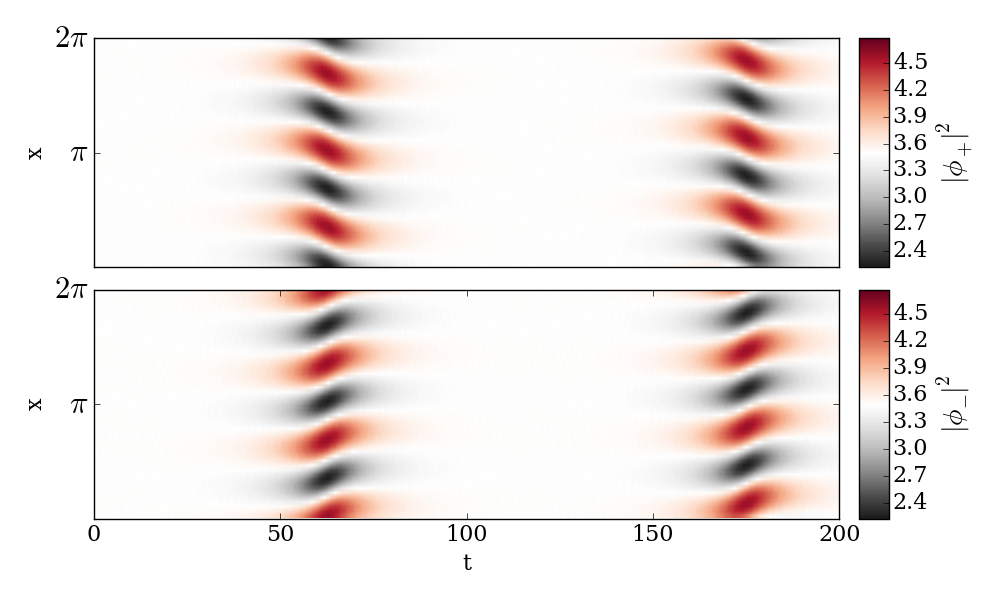}}
\put(3,0){(b)}
\end{picture}
&
\begin{picture}(35,22)
\put(0,0){\includegraphics[width=3.5in]{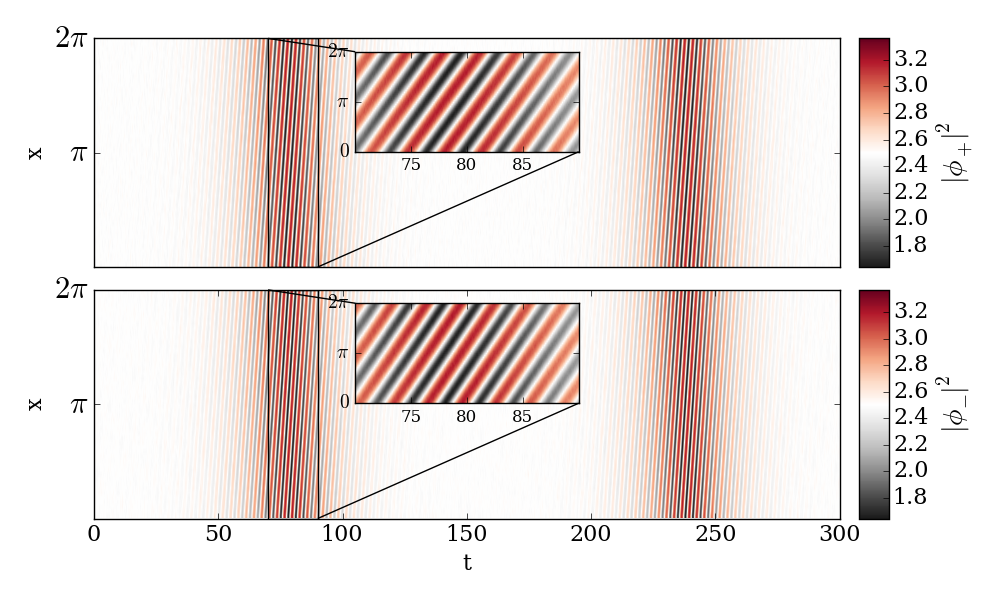}}
\put(3,0){(d)}
\end{picture}
\end{array}
$
\caption{\label{fig:instab-dynamics} 
(Color online)
Numerically calculated dynamics arising when an unstable states in the instability regions shown in Fig.~\ref{fig:z-rho} are distorted by a small initial perturbation. 
(a) $|l|=1$ instability in state $(-1,1)$ at $\rho=3$. (b) $|l|=3$ instability in state $(-2,2)$ at $\rho=7$. (c) $|l|=1$ and (d) $|l|=2$ instabilities in state $(m_+,m_-)=(-1,2)$ at $\rho=8$ and $\rho=5$, correspondingly. The onset of the modes with number of peaks equal to the angular momentum $|l|$ of the unstable mode is well visible, in agreement with Fig.~\ref{fig:z-rho}. 
In all cases $z=0$ for each state, $\alpha=-0.05$ and TE-TM splitting is absent $\kappa=0$. Video of the animated dynamics is available in the Supplementary Material.
}
\end{center}
\end{figure*}
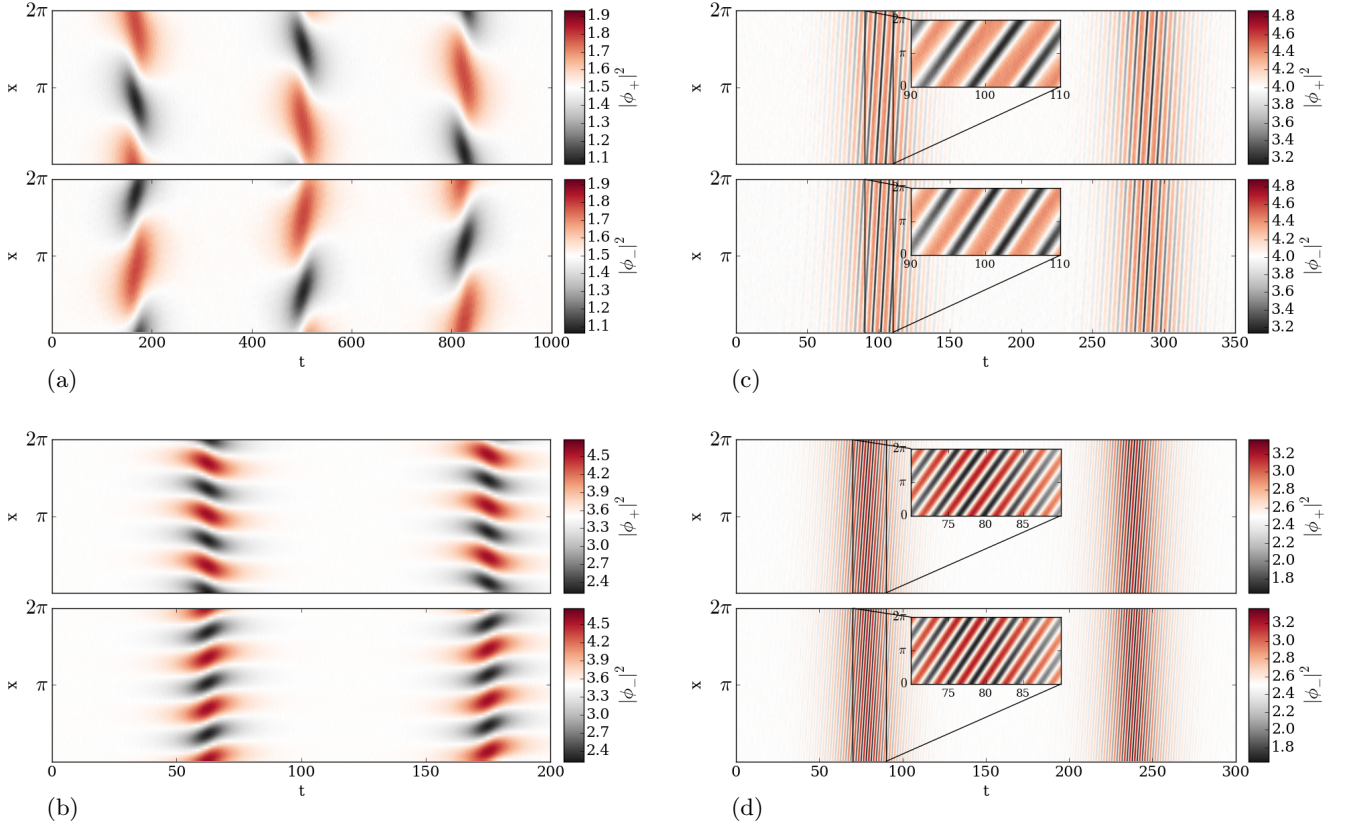
\fi

The other distinct class of solutions corresponds to the case when the both components have non-zero  densities,  $\chi_+,\chi_-\neq 0$. Equating the expressions in square brackets in~\eqref{chi-eq-general} and using the normalization $\chi_+^2+\chi_-^2=\rho$ we find
\begin{equation}
\chi_{+}^2=\frac{\rho-z}{2}, \quad
\chi_{-}^2= \frac{\rho+z}{2}
\label{chi-Meissner}
\end{equation}
where
\begin{equation}
z=\frac{2(\Omega-\Omega_{c})}{(1-\alpha)}
\label{z}
\end{equation}
and
\begin{equation}
\Omega_{c}=\frac12 (m_-^2-m_+^2)=0,\pm\frac12,\pm\frac32,\pm 2...
\label{Omc}
\end{equation}
Solutions~\eqref{chi-Meissner} depend on the magnetic field via parameter~$z=\chi_{-}^2-\chi_{+}^2$ subjected to the condition $|z|< \rho$. Therefore, solutions exist only in a limited interval of $\Omega$ given by
\begin{equation}
\left|\Omega-\Omega_{c}\right|<\rho\,\frac{(1-\alpha)}{2},
\label{condition-Meissner}
\end{equation}
and centered around $\Omega_c$. Note that $\Omega=\Omega_c$ ($z=0$) for different $m_+$, $m_-$ are degeneracy points at which energies~\eqref{mu-plus} and~\eqref{mu-minus} of the  circularly polarized solutions coincide. 

Substituting~\eqref{chi-Meissner} to~\eqref{chi-eq-general} we find chemical potential for the mixed polarization states~\eqref{chi-Meissner},
\begin{equation}
\mu^{(0)} = \frac12 \left[ m_+^2+m_-^2 + \rho (1+\alpha) \right].
\label{mu-Meissner}
\end{equation}
Thus, chemical potential $\mu=\mu^{(0)}$ in this case does not depend on the magnetic field and according to the terminology introduced in Section~\ref{sec:model}, these are the topological spin Meissner (TSM) states. 
Eqs.~\eqref{chi-Meissner},~\eqref{condition-Meissner} and~\eqref{mu-Meissner} are a generalization of the conventional spin Meissner effect to the case when spinor components possess non-trivial phase winding
and therefore can be called {\it topological} spin Meissner effect (TSM effect). 
The conventional spin Meissner effect~\cite{Rubo-PLA-2006, Shelykh-Superlattices, Larionov-PRL-2010} arises when spin-dependent polariton-polariton interactions compensate Zeeman splitting. Such compensation is possible until the fully circularly polarized state is reached. In the TSM effect Zeeman splitting in compensated by the combined action of both polariton-polariton interactions and circulation of the exciton-polariton condensate described by the winding numbers $m_+$, $m_-$. 
Note, that for TSM states to exists the offset of the magnetic field $\Omega$ from its critical values $\Omega_c$ should be small enough, see Eq.~\eqref{condition-Meissner}. 
The graphs of $\mu$ vs $\Omega$ and $\mu$ vs $\rho$ for the families of solutions~\eqref{mu-plus},~\eqref{mu-minus} and~\eqref{chi-Meissner} are shown in Figs.~\ref{fig:families}a and \ref{fig:families}b, respectively. The linear spectrum~\eqref{linear-zero} is recovered in the limit $\rho\to 0$, see Fig.~\ref{fig:families}b.

At a fixed magnetic field the condition~\eqref{condition-Meissner} defines the minimal value of the nonlinearity parameter $\rho$ which is needed to observe a given TSM state. Eqs. \eqref{Omc} and~\eqref{condition-Meissner} at $\Omega=0$ give the existence criterion  for a TSM state with the winding numbers~$m_+$ and~$m_-$,
\begin{equation}
\rho > \rho_{min}(m_+,m_-) \equiv \frac{|m_-^2 - m_+^2|}{1-\alpha}.
\label{rhomin}
\end{equation}
Therefore, more and more TSM states arise as nonlinearity is gradually increased from zero. The appearance of TSM states with increasing nonlinearity is seen on the  $\mu(\rho)$-plot on Fig.~\ref{fig:families}b and $\mu(\Omega)$-plot on Fig.~\ref{fig:largescale}. As seen from  Fig.~\ref{fig:families}b TSM states with $|m_+|=|m_-|$ start off straight from the linear spectrum ($\rho=0$) while those with $|m_+|\neq |m_-|$ require a finite value of nonlinearity given by~\eqref{rhomin}. In presence of large nonlinearity TSM states are the lowest energy states of the system and arrange into a system of energy levels as shown on Fig.~\ref{fig:largescale}.

An illustrative example of TSM effect is the behavior of half-vortices in the presence of magnetic field. Similar to half-vortices in a 2D system~\cite{Rubo-PRL-2007,Flayac-PRB-2010}, {\it half-vortices in a ring} have zero phase winding number of one component and a simple vortex in the other component. These are four distinct states $(1,0)$, $(-1,0)$, $(0,1)$ and $(0,-1)$. These are essentially nonlinear states which cease to exist in the linear limit, see Fig.~\ref{fig:families}b. In zero magnetic field half-vortices may only be observed for the nonlinearities stronger than the critical value $\rho > 1/(1-\alpha)$ given by the formula~\eqref{rhomin}. On the other hand, if magnetic field is tuned to $\Omega=\Omega_c=\pm 1/2$ even a very small nonlinearity would be enough to create a half-vortex. 

In order to compare our findings  with existing experimental studies of half-vortices in rings \cite{Liu-PNAS-2015} we change into the basis of linear polarization. At $z=0$ constant amplitude solution~\eqref{const-amp-sols} with amplitudes defined by~\eqref{chi-Meissner} takes the form
\begin{equation}
\psi_{\rm lin}(z=0) = \sqrt{\rho}\, \exp{\left[i \frac{m_++m_-}{2} x\right]}
\begin{pmatrix}
\cos{\frac{\Delta m}{2} x}\\
\sin{\frac{\Delta m}{2} x}\\
\end{pmatrix}
\label{hv-pure}
\end{equation}
where two components of the spinor in the basis of linear polarization are $\psi_{\rm lin,1}=(\psi_+ + i \psi_-)\sqrt{2}$ and $\psi_{\rm lin,2}=(\psi_+ - i \psi_-)/\sqrt{2}$.
Note, that the implicit choice of the relative phase of the spinor components $\chi_+=\chi_-$ made here is arbitrary: half-vortex states with different relative phases of $|\chi_+|$ and $|\chi_-|$ are connected by a simple shift of coordinate as will be shown in Section~\ref{section:sb}. For $\Omega$ away from $\Omega_c$ the expression for a half-vortex in linear polarization becomes more involved.
A simple expression may be obtained assuming $z/\rho \ll 1$,
\begin{equation}
\psi_{\rm lin} \approx \psi_{\rm lin}(z=0)
+i\frac{z}{\rho}
\begin{pmatrix}
\sin{\frac{\Delta m}{2} x}\\
-\cos{\frac{\Delta m}{2} x}\\
\end{pmatrix}
\label{psihv}
\end{equation}
The formula~\eqref{psihv} can explain the experimental result~\cite{Liu-PNAS-2015} where a superposition of half-vortex states was observed. According to Ref.~\cite{Liu-PNAS-2015} the anzatz in the form of superposition of states of the type~\eqref{hv-pure}  was used to fit the experimental data.
This puzzling result could not be explained by the existing theory but created an uncertainty of why half-vortices prefer a superposition in favor of a pure state~\eqref{hv-pure}. Assuming the experiments were done in zero magnetic field we find that $|z|={1/(1-\alpha)\neq 0}$. Therefore, formula~\eqref{psihv} makes it clear that a pure half-vortex states in the form~\eqref{hv-pure} can not be observed in zero magnetic field. Furthermore, if a pure half-vortex state~\eqref{hv-pure} is to be observed, a non-zero magnetic field with magnitude equal exactly to $|\Omega|=1/2$ should be applied. As far as we know, this has not been done in the existing experimental studies of exciton-polariton states in a ring.

\subsection{Stability of spin Meissner states}

To analyze stability we consider small time-dependent perturbations $\varepsilon_{\pm}(x,t)$ around vortices:
\begin{equation}
\psi_{\pm}(x,t)=\left[\chi_{\pm}+\varepsilon_{\pm}(x,t)\right]e^{i m_{\pm} x}.
\label{psi-eps}
\end{equation}
Substituting~\eqref{psi-eps} into~\eqref{GP-ring} at $\kappa=0$ we get a system of linear equations for $\varepsilon_{\pm}(x,t)$,
\begin{multline}
i\dot\varepsilon = -\varepsilon_{\pm}'' - 2i m_{\pm} \varepsilon_{\pm}' 
 + \chi_{\pm}^2(\varepsilon_{\pm}+\varepsilon_{\pm}^*)+
\alpha\chi_{\pm}\chi_{\mp}(\varepsilon_{\mp}+\varepsilon_{\mp}^*) =0
\end{multline}
Expanding $\varepsilon_{\pm}(x,t)$ into Fourier series in $x$, see, e.g., Ref. \cite{Skryabin-PRA-2000},
\begin{equation}
\varepsilon_{\pm}(x,t) = \sum_{l=0}^{\infty} U_{\pm,l}(t) e^{ilx} + V_{\pm,l}^*(t) e^{-ilx}
\label{eps-Fourier}
\end{equation}
we get a set systems of equations on $\mathbf{W}_l(t)=(U_{+,l},V_{+,l},U_{-,l},V_{-,l})$, decoupled for different integer~$l$,
\begin{equation}
i\dot{\mathbf{W}_l} = \hat{\eta}\hat{\mathcal{H}_l}\mathbf{W}_l
\end{equation}
where 
\begin{equation}
\hat\eta=
\begin{pmatrix}
1 & 0 & 0 & 0 \\
0 & -1 & 0 & 0 \\
0 & 0 & 1 & 0 \\
0 & 0 & 0 & -1
\end{pmatrix}
\label{eta}
\end{equation}
and
\begin{equation}
\hat{\mathcal{H}_l}=\begin{pmatrix}
d_+ & \chi_+^2 & \alpha\chi_+\chi_- & \alpha\chi_+\chi_-
\\
\chi_+^2 & \tilde{d}_+ & \alpha\chi_+\chi_- & \alpha\chi_+\chi_-
\\
\alpha\chi_+\chi_-  & \alpha\chi_+\chi_- & d_- & \chi_-^2
\\
\alpha\chi_+\chi_- & \alpha\chi_+\chi_-  & \chi_-^2 & \tilde{d}_-
\end{pmatrix}
\label{H0}
\end{equation}
where $
d_\pm \equiv 
l^2 + 2l m_\pm + \chi_{\pm}^2$, $\tilde{d}_\pm \equiv 
l^2 - 2l m_\pm + \chi_{\pm}^2$.
Assuming $\mathbf{W}_l= \mathbf{w}_l e^{-i\lambda t}$, $\mathbf{w}_l\equiv (u_{+,l},v_{+,l},u_{-,l},v_{-,l})$  we get an eigenvalue problem:
\begin{equation}
\hat\eta \hat{\mathcal{H}_l}\mathbf{w}_l=\lambda \mathbf{w}_l
\label{etaH-problem}
\end{equation}
The solution is spectrally unstable if there is at least one eigenvalue with positive imaginary part $\Im\lambda > 0$. Because of equality ${\rm Tr}[(\hat{\eta}\hat{\mathcal{H}_l})^\dagger]={\rm Tr}[\hat{\eta}\hat{\mathcal{H}_l}]$ the eigenvalues of the matrix $\hat{\eta}\hat{\mathcal{H}_l}$ come in complex conjugated pairs.

In the special case $\Omega=\Omega_{c}$ the eigenvalue problem~\eqref{etaH-problem},~\eqref{H0} allows simple analytical solution. For $\Omega=\Omega_{c}$ we have, according to~\eqref{chi-Meissner}, $\chi_+=\chi_-=\sqrt{\rho/2}$. Substituting to~\eqref{etaH-problem},~\eqref{H0} and solving the eigenvalue problem we get four eigenvalues 
\begin{multline}
\lambda= l \Big( m_+ + m_- 
\\
\pm \sqrt{\Delta m^2+l^2+\rho\pm
\sqrt{4\Delta m^2(l^2+\rho)+\alpha^2\rho^2}}\,\Big)
\label{lambdas-z0}
\end{multline}
where 
\begin{equation}
\Delta m \equiv m_- - m_+
\end{equation}
From~\eqref{lambdas-z0}, the unstable regions can be easily found,
\begin{equation}
|l^2 + \rho - \Delta m^2| < \alpha\rho
\label{instab-at-Omc}
\end{equation}
It is straightforward to see from~\eqref{instab-at-Omc} that there is no instability in the case $\alpha=0$ when interaction between the circular components is absent.
However, for  finite $\alpha$ we have an unstable region in $\rho$
extending up to $\Delta\rho\approx |\alpha|\rho_c$ either side from  $\rho_{c}= \Delta m^2-l^2$.

We analyze stability for arbitrary values of $\Omega$ we use the perturbation theory in parameter $\alpha$. 
We take into account the dependence of $\chi_+$ and $\chi_-$ on $\alpha$ exactly using their expressions~\eqref{chi-Meissner}, while we treat perturbatively only those part of~\eqref{H0} depends on $\alpha$ explicitly.
\begin{equation}
\mathcal{H}_l=\mathcal{U}_l+\alpha \mathcal{V}(\alpha)
\end{equation}
\begin{equation*}
\mathcal{U}_l=\begin{pmatrix}
d_+ & \chi_+^2 & 0 & 0
\\
\chi_+^2 & \tilde{d}_+ & 0 & 0
\\
0  & 0 & d_- & \chi_-^2
\\
0 & 0  & \chi_-^2 & \tilde{d}_-
\end{pmatrix}
,\;\;
\mathcal{V}(\alpha)=\chi_+\chi_-\begin{pmatrix}
0 & 0 & 1 & 1
\\
0 & 0& 1 & 1
\\
1  & 1 & 0 & 0
\\
1 & 1  & 0 & 0
\end{pmatrix}
\end{equation*}
For eigenvalues of matrix $\eta\mathcal{U}_l$ we find simple expressions
\begin{equation}
\begin{split}
\lambda_{1,2}^{+} = l\left[2m_+ \pm \sqrt{l^2+2\chi_{+}^2}\right], \\
\lambda_{1,2}^{-} = l\left[2m_- \pm \sqrt{l^2+2\chi_{-}^2}\right].
\end{split}
\label{lambdas}
\end{equation}
Eigenvalues of $\eta\mathcal{H}$ with non-zero imaginary part may appear around degeneracy points of $\eta\mathcal{U}_l$. In the trivial case $\Delta m=0$ the degeneracies may only appear when $\chi_-^2=\chi_+^2$, i.e. at $z=0$ which is the case studied above: the eigenvalues of matrix~\eqref{H0} are given by~\eqref{lambdas-z0}. As seen from Eq.~\eqref{lambdas-z0} this does not lead to any instabilities as soon as $|\alpha|\le 1$. 
In follows we will assume $\Delta m\neq 0$. Equating the eigenvalues $\lambda_{1,2}^{+}$ and $\lambda_{1,2}^{-}$ we get for degeneracies  $\lambda^{+}_1=\lambda^{-}_2$ ($\Delta m >0$) and $\lambda^{+}_2=\lambda^{-}_1$ ($\Delta m <0$) realized when 
$|z|\le 2\Delta m^2$,
\begin{equation}
\rho_c(z) = \Delta m^2 - l^2 + \frac{z^2}{4\Delta m^2}
\label{rhoc}
\end{equation}
Applying the perturbation theory for non-Hermitian operators~\cite{Sternheim-1972} we find the first order corrections to the degenerate eigenvalues. At $\rho=\rho_c$ these are
\begin{equation}
\Delta \lambda = \pm \frac{i \alpha l \chi_+ \chi_- }{\left[(l^2+2\chi_+^2)(l^2+2\chi_-^2)\right]^{1/4}}
\label{dlambda}
\end{equation}
and for $l\neq 0$ lead to instability due to appearance of an eigenvalue with positive imaginary part.

Treating the deviation $\rho-\rho_c$ from $\rho_c$ as a perturbation we find the size of the instability interval $(-\Delta\rho,\Delta\rho)$ centered around $\rho_c$ given by the formula
\begin{equation}
\Delta\rho = \left(  \frac{ 8 \alpha^2 \chi_+^2\chi_-^2 \sqrt{l^2+2\chi_+^2}\sqrt{l^2+2\chi_-^2}}{l^2+\chi_+^2+\chi_-^2 + \sqrt{l^2+2\chi_+^2} \sqrt{l^2+2\chi_-^2}} \right)^{1/2}
\label{Delta-rho}
\end{equation}
where $\chi_+$, $\chi_-$ are evaluated using Eq.~\eqref{chi-Meissner} at $\rho=\rho_c$.

One may check that the other two degeneracies, ${\lambda^{+}_1=\lambda^{-}_1}$ and $\lambda^{+}_2=\lambda^{-}_2$ which realize when $|z|\ge 2\Delta m^2$ do not lead to imaginary eigenvalues and, therefore, do not cause instabilities.

Formula~\eqref{rhoc} together with condition $|z|<\rho$ imply that the four half-vortices $(1,0)$, $(-1,0)$, $(0,1)$ and $(0,-1)$ have no instability regions, i.e. linearly stable for all values of $\rho$. The first non-trivial case arises for states with $|\Delta m|=2$. Instability regions for states with $|\Delta m|=2, 3, 4$ and $10$ are shown in a parameter space of $\rho$ vs $z$ in Fig.~\ref{fig:z-rho}.

We estimate the dynamical effect of the unstable mode on distribution of the spinor components. Assuming the unstable dominates other modes but still can be treated as a perturbation around the stationary solution, we write
\begin{equation}
\varepsilon_{\pm} \approx \left[ u_{\pm} e^{ilx-i\lambda_r t} + v^*_{\pm} e^{-ilx+i\lambda_r t}\right] e^{\lambda_i t}
\label{eps-unstable}
\end{equation}
where we explicitly separated the real $\lambda_r =\Re \lambda$ and imaginary $\lambda_i =\Im \lambda$ parts of the eigenvalue causing the instability. 
Substituting~\eqref{eps-unstable} to the expressions for densities 
$|\phi_{\pm}|^2 = |\chi_{\pm}+\varepsilon_{\pm}(x,t)|^2$
we see that the growing unstable mode modulates densities of the spinor components in the form of a propagating wave with phase velocity $v=\lambda_r/l$.
The real part $\lambda_r$ can be estimated from expressions~\eqref{lambdas}. At degeneracies $\lambda^{+}_1=\lambda^{-}_2 \equiv \lambda$ and $\lambda^{+}_2=\lambda^{-}_1 \equiv \lambda$ we find for the phase velocity at $\rho= \rho_c(z)$,
\begin{equation}
v = m_+ + m_- - \frac{z}{2\Delta m}
\label{v}
\end{equation}
Thus, $v$ is determined by the total angular momentum $m_+ + m_-$ of a TSM state.

We use the Split Step (Fourier) method to numerically analyze dynamics of the unstable states. The $(-1,1)$, $(-2,2)$ and $(-1,2)$ cases are shown in Fig.~\ref{fig:instab-dynamics}. 
The onset of unstable mode with number of peaks or deeps is equal to the angular harmonics $l$ in agreement with the corresponding instability regions in Fig.~\ref{fig:z-rho}.
The instability develops as a wave of density modulations. The unstable modes with $l=1$ and $l=2$ around the state $(-1,2)$ cause the density modulation to rotate with phase velocity $v\approx 1$, see Figs.~\ref{fig:z-rho}c,d. While density modulations in state $(-1,2)$ move anticlockwise (increasing $x$),  density modulations in state $(-2,1)$ moves clockwise in agreement with the opposite sign of the phase velocity in~Eq.~\eqref{v} (see Supplementary Material). 

\section{Constant-amplitude spin Meissner states in presence of TE-TM splitting.}
\label{section:const-amp}

\iffigures
\setlength{\unitlength}{0.1in}
\begin{figure}
\begin{center}
$
\begin{array}{c}
\begin{picture}(35,23)
\put(-1,0){\includegraphics[width=3.4in]{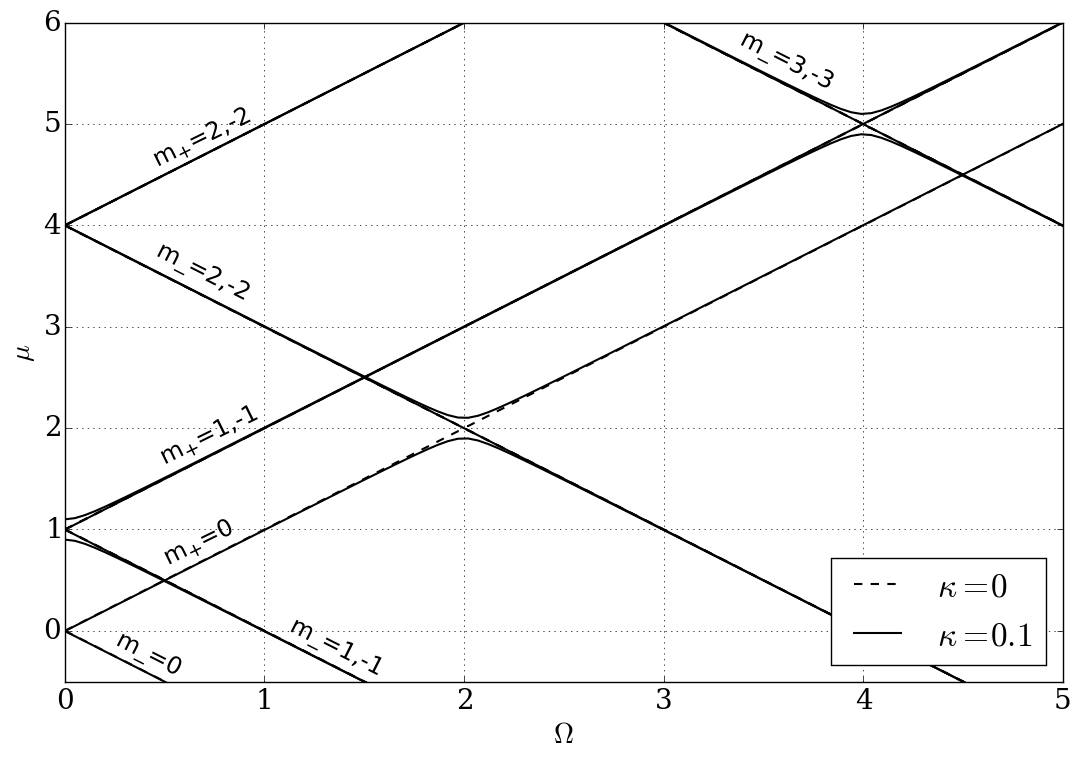}}
\end{picture}
\end{array}
$
\caption{\label{fig:linear} 
Linear spectrum in presence (solid lines) and absence (dashed lines) of TE-TM splitting. TE-TM splitting results in anticrossings at $\Omega=\Omega_c\equiv 2n$, $n=0,\pm 1,...$
for branches $m_\pm=n\pm 1$, see Eq.~\eqref{linear-kappa}.
}
\end{center}
\end{figure}
\fi

\iffigures
\begin{figure*}
\begin{center}
\includegraphics[width=7in]{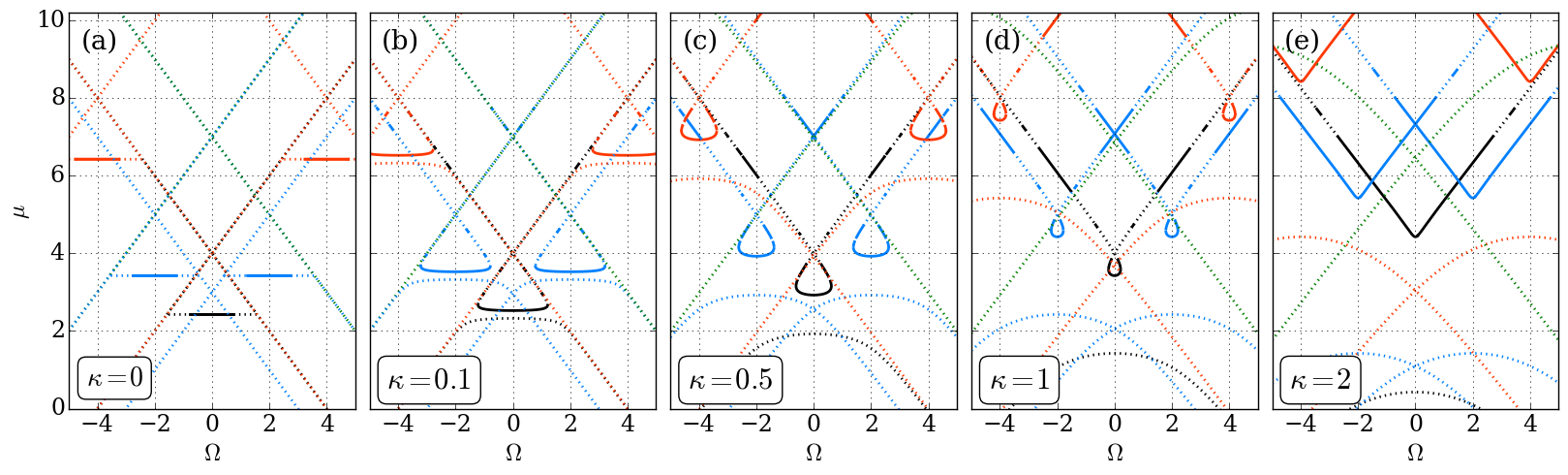}
\caption{\label{fig:rho3} 
(Color online) 
The graphs show evolution of constant-amplitude solutions of the system~\eqref{GP-ring} with increasing TE-TM splitting for different values of the phase winding number $m$. Black, blue, red and green lines correspond to the states with $|n|=0,1,2$ and $3$. TE-TM splitting parameter~$\kappa$ is increased from 0.0 (a) to 0.1 (b), 0.5 (c), 1.0 (d) and 2.0 (e). The nonlinearity parameters are $\rho=3$ and $\alpha=-0.05$. Splitting of constant amplitude TSM states into stable and unstable branches is visible on Fig.b when non-zero $\kappa$ is introduced. For visibility purposes unstable regions are marked by solid lines and stable regions are marked by dotted lines. Note that only constant amplitude solutions are shown here. 
}
\end{center}
\end{figure*}
\fi

\iffigures
\setlength{\unitlength}{0.1in}
\begin{figure}
\begin{center}
$
\begin{array}{c}
\begin{picture}(35,23)
\put(-1,0){\includegraphics[width=3.4in]{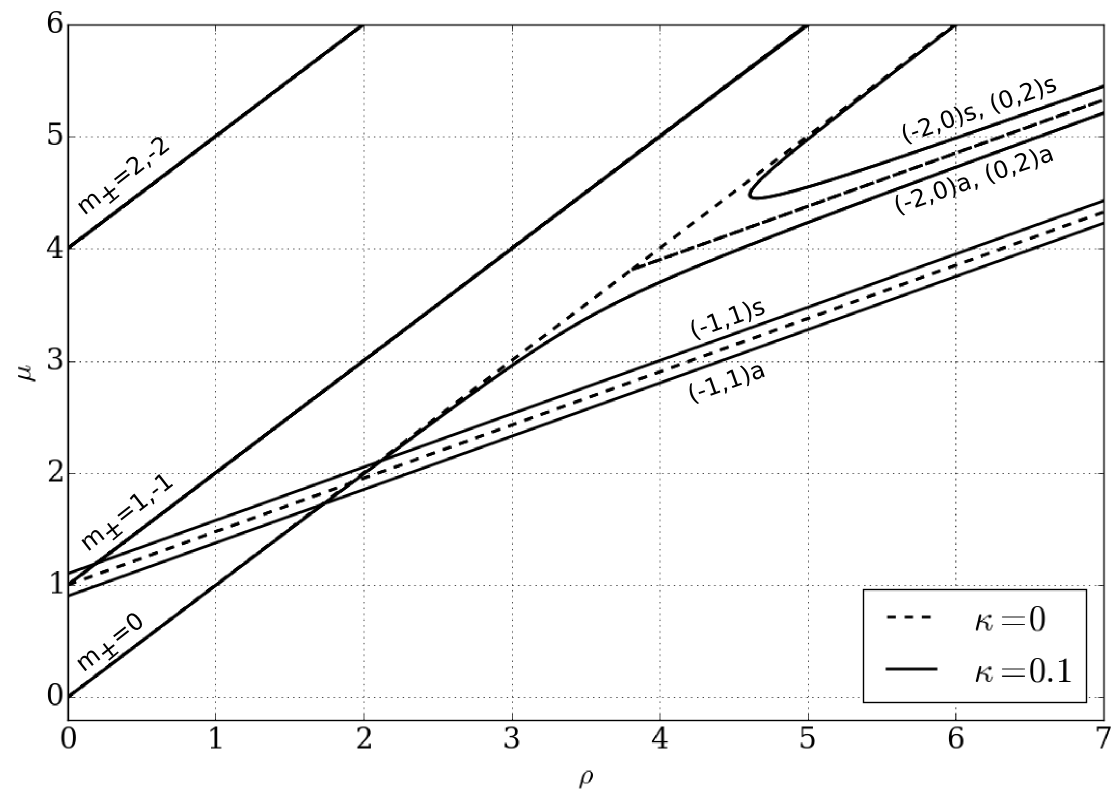}}
\end{picture}
\end{array}
$
\caption{\label{fig:rho-mu} 
Splitting of the spin Meissner states $\Delta m=2$ under a non-zero $\kappa$: states $(-1,1)$, $(-2,0)$ and $(0,2)$ shown here split into symmetric ($\chi_+\chi_->0$) and antisymmetric ($\chi_+\chi_-<0$) branches marked by `s' and `a' after the brackets with a pair of winding numbers $(m_+,m_-)$. Dashed line and solid lines correspond to $\kappa=0$ and $\kappa=0.1$, respectively. Other parameters are $\alpha=-0.05$ and $\Omega=0$. For simplicity, only constant amplitude solutions are shown here.
}
\end{center}
\end{figure}
\fi

\iffigures
\setlength{\unitlength}{0.1in}
\begin{figure}
\begin{center}
$
\begin{array}{cc}
\begin{picture}(17,20)
\put(-1,0){\includegraphics[width=1.75in]{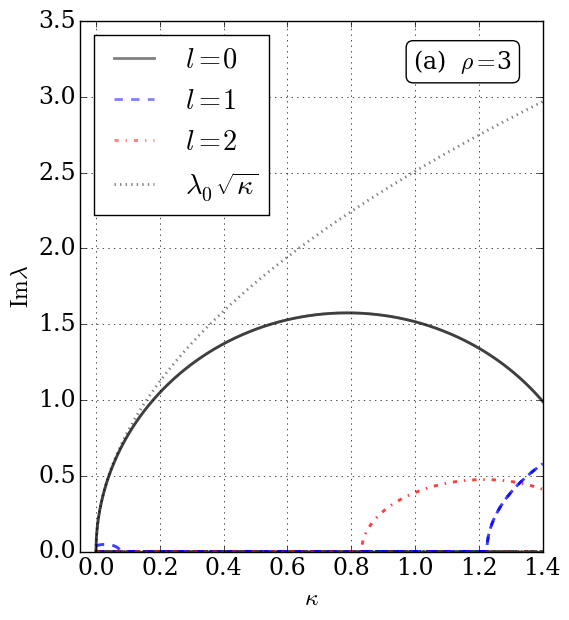}}
\end{picture}
& 
\begin{picture}(17,20)
\put(-1,0){\includegraphics[width=1.75in]{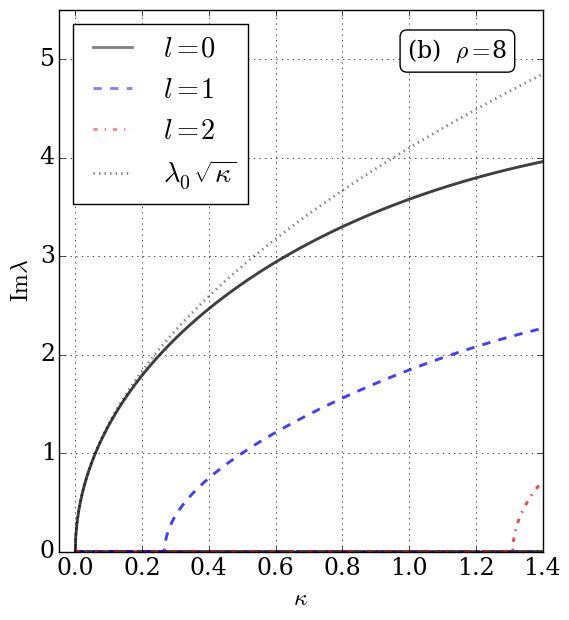}}
\end{picture}
\end{array}
$
\caption{\label{fig:lambda-unstable} 
(Color online)
Imaginary part of the eigenvalues of matrix~\eqref{H-with-kappa} as a function of $\kappa$, evaluated at the unstable branch of the state $(-1,1)$ and $\Omega=0$. Black, dashed and dashed-dotted lines correspond to instabilities caused by angular harmonics $l=0$, $1$ and $2$, respectively. Dotted line is the theoretical estimate given by the first term in formula~\eqref{lambda-expansion} with~\eqref{lambda02}.
}
\end{center}
\end{figure}
\fi

\iffigures
\setlength{\unitlength}{0.16in}
\begin{figure}
\begin{center}
\includegraphics[width=3.3in]{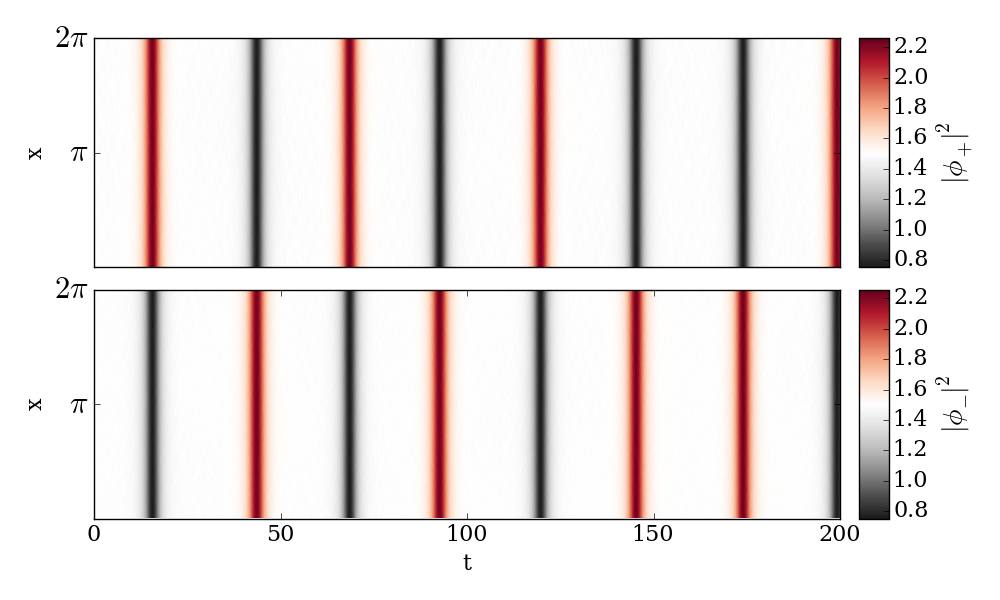}
\caption{\label{fig:stability-1} 
(Color online) 
Numerically calculated dynamics of instability arising when unstable state $(-1,1)$ at $\rho=3$, $\Omega=0$ is distorted by a small initial perturbation in presence of TE-TM splitting $\kappa=0.1$. The onset of $|l|=0$ mode is visible in agreement with Figure (a). Video of the animated dynamics is available in the Supplementary Material.
}
\end{center}
\end{figure}
\fi

We now going to look at how the presence of a non-zero TE-TM splitting 
influences TSM states and the topological spin Meissner effect.

It is convenient to eliminate the explicit dependence on $x$ from the TE-TM splitting term, which is achieved through the substitution 
\begin{equation}
\psi_{\pm} = 
\phi_{\pm}(x) e^{\mp i x}.
\end{equation}
With this substitution we get the following auxiliary system of equations which takes a  rotationally invariant form
\begin{eqnarray}
\nonumber && i\dot\phi_+ = \left[ \hat D_{+} +|\phi_+|^2 + \alpha |\phi_-|^2 \right] \phi_+    + \kappa \phi_-
\\
&& i\dot\phi_- = \left[ \hat D_{-} +|\phi_-|^2 + \alpha |\phi_+|^2 \right] \phi_-    + \kappa \phi_+
\label{GP-phi}
\end{eqnarray}
where
$\hat D_{\pm}=1-\mu-\p_x^2 \pm 2i\p_x \pm \Omega$. The form of equation~\eqref{GP-phi} is explicitly invariant under rotations, i.e. a shift of coordinate $\mathcal{\hat R}(\Delta x)\phi_\pm(x)=\phi_\pm(x+\Delta x)$. 

In the linear regime and $\kappa\neq 0$, the stationary version of the system~\eqref{GP-phi} is a linear system of equation with constant coefficients which has solutions in the form of exponentials 
\begin{equation}
\phi_{\pm}(x)=\chi_{\pm}e^{inx},~\kappa\ne 0 
\label{const-amp-n}
\end{equation}
with the winding number $n$ for both components and amplitudes
\begin{equation}
\chi_+ = \sqrt{\frac{\rho}{1+\xi^2}}, \quad
\chi_- = \xi\,\sqrt{\frac{\rho}{1+\xi^2}}
\label{lin-sol-kappa}
\end{equation}
where $\xi\equiv =\chi_-/\chi_+$ is given by
\begin{equation}
\xi=\xi_\pm,\quad \xi_\pm \equiv \frac{2n - \Omega}{\kappa} \pm \sqrt{1+\left(\frac{2n - \Omega}{\kappa}\right)^2}
\end{equation}
Energies of the solutions~\eqref{lin-sol-kappa} are
\begin{equation}
\mu = 1+n^2 \pm \sqrt{(\Omega-2n)^2+\kappa^2}.
\label{spectrum-linear}
\end{equation}
Note, that the winding numbers in the $\psi_{\pm}$ representation (Eq.~\eqref{GP-ring}) are given by
\begin{equation}
m_{\pm}=n\mp 1, \quad \Delta m=2
\end{equation}
The linear spectrum~\eqref{spectrum-linear} is plotted on Figs.~\ref{fig:linear} for the cases of zero ($\kappa=0$) and non-zero ($\kappa=0.1$) TE-TM splitting. The splitting of energy levels at $\Omega=\Omega_c\equiv 2 n$ caused by TE-TM splitting is seen on Fig.~\ref{fig:linear}b. The avoided crossings arrange in a parabolic pattern $\mu(\Omega_c)\sim \Omega_c^2$ as given by the formula~\eqref{spectrum-linear}. 

We now look into the nonlinear case. Constant amplitude solutions of the nonlinear system~\eqref{GP-phi} have the same form~\eqref{const-amp-n}
where the amplitudes~\eqref{lin-sol-kappa} are defined by real roots of the 4th order algebraic equation on $\xi$,
\begin{multline}
\kappa(\xi^4-1)-\rho\left(1-\alpha\right)(\xi^3-\xi) \\
+2(\Omega-2n)(\xi^3+\xi)=0
\label{ring-sol-nln}
\end{multline}
which, in general, may have 4, 2 or 0 real roots.
The bifurcations between pairs of real and complex roots when changing the magnetic field and 
strength of TE-TM splitting $\kappa$ can be traced on Fig.~\ref{fig:rho3}. The figure shows evolution of the constant amplitude branches with changing strength of TE-TM splitting $\kappa$ at a fixed nonlinearity $\rho=3$. The TSM branches exist for small $\kappa$ keeping their magnetic field-independent form of chemical potential. On increasing $\kappa$, the topological spin Meissner effect in the lower branch gradually comes to a naught acquiring a parabolic dependence on $\Omega$, while the top branch disappears completely at high $\kappa$.

Although, the exact solutions can be found by solving the equation~\eqref{ring-sol-nln} it is instructive to find their explicit expressions in the limit of small $\kappa$ (see Supplementary Information). In the first order in $\kappa$ we find 
for TSM states $\Delta n =0$,
\begin{equation}
\mu = \mu^{(0)} + \frac{\kappa\,\rho}{2\chi_+\chi_-} + O(\kappa^2)
\end{equation}
where zero order term $\mu^{(0)}$ is given by~\eqref{mu-Meissner}
and ${\chi_+\chi_->0}$, ${\chi_+\chi_-<0}$ are two distinct branches of solutions. Splitting of the TSM states $\Delta m =2$ into symmetric ($\chi_+\chi_->0$) and antisymmetric ($\chi_+\chi_-<0$) when a non-zero $\kappa$ is introduced, is shown on the Fig.~\ref{fig:rho-mu} 
In zero magnetic field and absence of interactions, $(-1,1)s$ and $(-1,1)a$ are two lowest TE and TM modes in a ring, while higher order states such as $(-2,0)s$, $(0,2)s$ and $(-2,0)a$ and $(0,2)a$ are propagating TE and TM modes with non-zero wavevector.
In the linear limit $\rho\to 0$ the splitting of the $(-1,1)$ state is the avoided crossings given by linear spectrum~\eqref{spectrum-linear} and shown on Fig.~\ref{fig:linear}.


We analyze analytically stability of constant amplitude TSM states solving perturbatively the eigenvalue problem for operator $\hat L = \eta \hat H$, where operator $\hat H$ is given in Appendix~\ref{app:stab-th}. Because the constant amplitude TSM states are split in two branches when non-zero $\kappa$ is present, perturbation expansion for eigenvalues of the operator 
$\hat L=\hat L_0+\kappa\hat L_1+O(\ka^2)$ will involve powers of $\kappa^{1/2}$,
\begin{equation}
\la=\ka^{1/2}\left[ \la_0 +O(\kappa)\right]
\label{lambda-expansion}
\end{equation}
where for $\lambda_0$ we find (see details in Appendix~\ref{app:stab-th}) 
\begin{equation}
\lambda_0^2 = \mp 2(1-\alpha)\sqrt{\rho^2-z^2}
\label{lambda02}
\end{equation}
for states $\chi_+ \chi_- >0$ and $\chi_+\chi_-<0$, correspondingly. Therefore, for $\kappa>0$ the state with $\chi_+=\chi_-$ is unstable and state $\chi_+=-\chi_-$ is stable (for $\kappa<0$ the situation reverses).
For the unstable mode with eigenvalue given by~\eqref{lambda02} we find
\begin{equation}
\varepsilon_{\pm} \sim \pm \chi_\pm e^{inx} = \pm \phi_{\pm}(x)
\label{l0-mode}
\end{equation}
i.e. the unstable mode~\eqref{l0-mode} is homogeneous over the ring.

We compare the theoretical estimate~\eqref{lambda-expansion}, ~\eqref{lambda02} of the imaginary part of the eigenvalue $\lambda$ causing the instability to its exact value obtained by numerical diagonalization of matrix~\eqref{H-with-kappa} for different angular harmonics $l$. The results of comparison for dependence of ${\rm Im}\lambda$ on $\kappa$ is presented in  Figs.~\ref{fig:lambda-unstable}a,b for two different values of nonlinearity parameter $\rho$. 
The theoretical results~\eqref{lambda-expansion}, ~\eqref{lambda02} obtained by perturbation expansion agree with numerical calculation for $l=0$ at small $\kappa$.
 A small region of instability caused by angular harmonics $l=1$ is also visible in Figure~\ref{fig:lambda-unstable}a for small values of $\kappa$ and is caused by the instability region $l=1$ on Fig.~\ref{fig:z-rho}a (see Fig.~\ref{fig:z-rho}a at $\rho=3$). With increasing strength of TE-TM splitting a state becomes unstable with respect to several harmonics simultaneously.

In our numerical analysis of stability we solve the eigenvalues problem of the operator $\hat{\eta}\hat{\mathcal{H}}_l(\kappa)$, where (see Appendix~\ref{app:stab-num} for details)
\begin{multline}
\hat{\mathcal{H}}_l(\kappa)=
\\
\begin{pmatrix}
d_+ & \chi_+^2 & \alpha\chi_+\chi_- +\kappa & \alpha\chi_+\chi_-
\\
\chi_+^2 & \tilde{d}_+ & \alpha\chi_+\chi_- & \alpha\chi_+\chi_- +\kappa
\\
\alpha\chi_+\chi_- +\kappa & \alpha\chi_+\chi_- & d_- & \chi_-^2
\\
\alpha\chi_+\chi_- & \alpha\chi_+\chi_- +\kappa & \chi_-^2 & \tilde{d}_-
\end{pmatrix},
\label{H-with-kappa}
\end{multline}
$\hat\eta$ was defined above and the diagonal elements $d_{\pm}, \tilde{d}_{\pm}$ are given by
\begin{multline}
d_\pm \equiv 
l^2 + 2l m_{\pm} + \chi_{\pm}^2 \\
+ \left[-\mu+m_{\pm}^2+ \chi_{\pm}^2+\alpha\chi_{\mp}^2 \pm\Omega\right]
\end{multline}
\begin{multline}
\tilde{d}_\pm \equiv 
l^2 - 2l m_{\pm} + \chi_{\pm}^2 \\
+ \left[-\mu+m_{\pm}^2 + \chi_{\pm}^2+\alpha\chi_{\mp}^2 \pm\Omega\right].
\end{multline}
Matrix \eqref{H-with-kappa} is a generalization of~\eqref{H0} to nonzero $\kappa$ 

The solution is spectrally unstable if there is at least one eigenvalue with positive imaginary part $\Im\lambda > 0$. The results of the stability analysis are shown on Figure~\ref{fig:rho3}. Dashed lines mark stable regions and solid lines mark unstable regions (stability analysis with respect to the individual harmonics $l$ can be found in the Supplementary Material). 
As seen from Fig.~\ref{fig:rho3}, the constant amplitude TSM states are split into 
a stable (bottom) and unstable (top) branches, in agreement with~\eqref{lambda02}.
At $\kappa=0$ the instability is caused by angular harmonics $l=1$ which corresponds to the instability region shown in Fig.~\ref{fig:z-rho}a while for non-zero $\kappa$ the $l=0$ mode appears as seen in Fig.~\ref{fig:lambda-unstable}a.
 
Finally, we analyze dynamics of instability arising due to the unstable mode in presence of TE-TM splitting. The results of our time-dependent numerical calculations are shown on Fig.~\ref{fig:stability-1}. As seen from the Figure presence of TE-TM splitting leads to a homogeneous instability mode in agreement with formula~\eqref{l0-mode}.

\section{Symmetry breaking spin Meissner states}
\label{section:sb}

\iffigures
\setlength{\unitlength}{0.1in}
\begin{figure}
\begin{center}
$
\begin{array}{c}
\begin{picture}(35,23)
\put(-1,0){\includegraphics[width=3.4in]{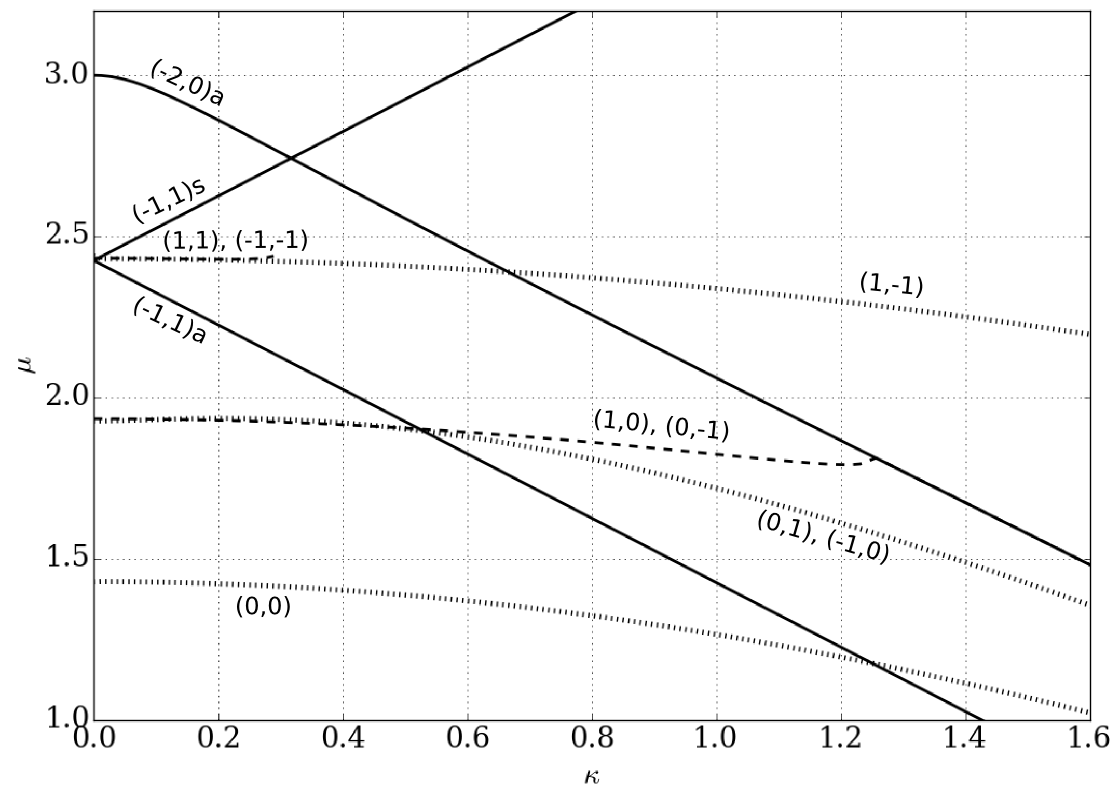}}
\end{picture}
\end{array}
$
\caption{\label{fig:kappa-mu-rho3} 
Numerical continuation of states in parameter $\kappa$ to the region of non-zero TE-TM splitting. Solid lines mark constant amplitude solutions, dashed and dotted lines mark symmetry-breaking states. Splitting of the state $(-1,1)$ into two branches can be seen. Contrary to the constant amplitude states, symmetry-breaking states do not split into branches but develop inhomogeneous density profiles. The $(1,1)$, $(-1,-1)$ develop instabilities at $\kappa\approx 0.3$  which lead to breaking of the branch. Other parameters are $\alpha=-0.05$ and $\Omega=0$.
}
\end{center}
\end{figure}
\fi

\iffigures
\setlength{\unitlength}{0.1in}
\begin{figure}
\begin{center}
\includegraphics[width=3.4in]{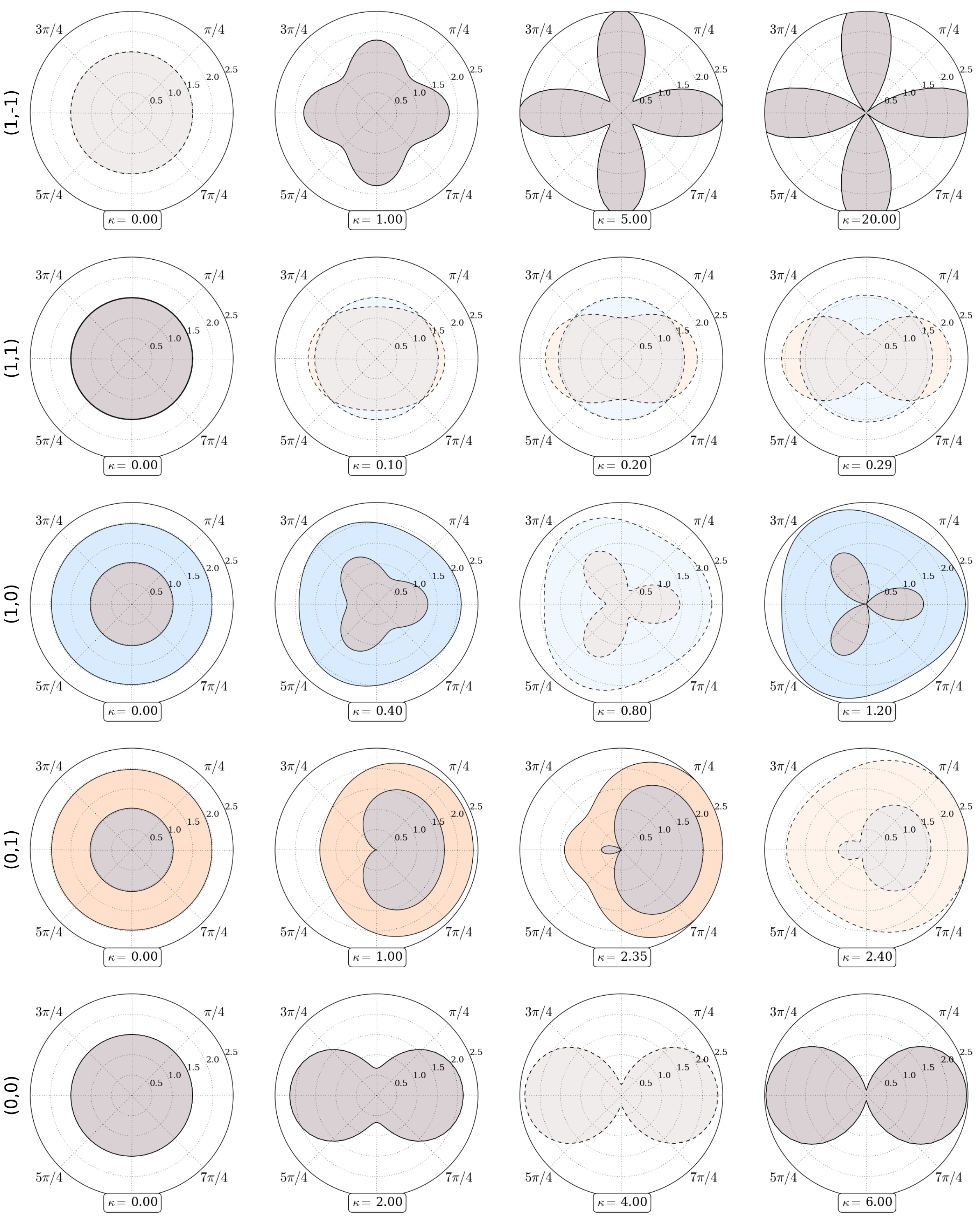}
\caption{\label{fig:images-rho3} 
(Color online) 
Numerical continuation of TSM states given by Eqs.~\eqref{chi-Meissner} at $\kappa=0$ to the domain of non-zero TE-TM splitting. Evolution of the densities 
of the spinor components of symmetry-breaking states during the continuation is shown with parameter $\kappa$ increasing from left to right. Polar plots correspond to the branches of the symmetry-breaking states on Fig.~\ref{fig:kappa-mu-rho3} with the same nonlinearity strength $\rho=3$. Parameters of the calculations are $\Omega=0$ and $\alpha=-0.05$. States $(-1,0)$ and $(0,-1)$ are not shown as they evolve in the same way as $(0,1)$ and $(1,0)$, respectively, with spinor components interchanged. Dashed lines and fainter colors mark unstable states.
}
\end{center}
\end{figure}
\fi

\iffigures
\setlength{\unitlength}{0.1in}
\begin{figure}
\begin{center}
\includegraphics[width=3.4in]{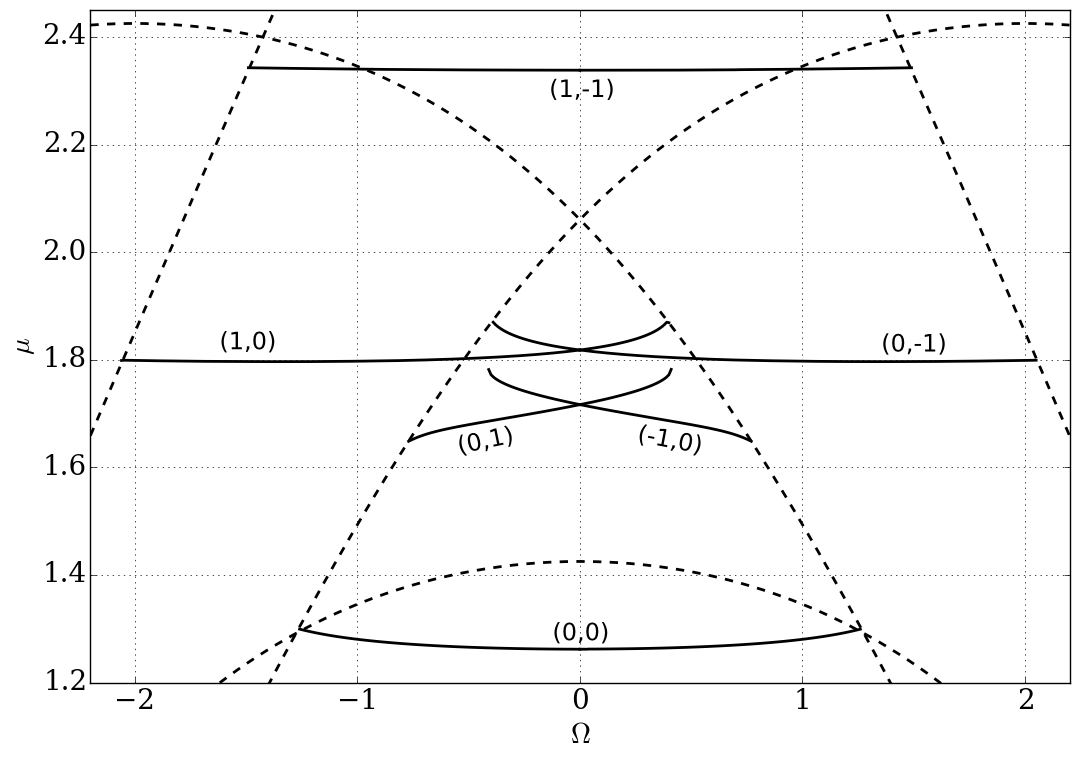}
\caption{\label{fig:Omega-mu-rho3} 
(Color online) 
Topological spin Meissner effect in presence of TE-TM splitting.
As seen from the Figure energies of symmetry-breaking TSM states (solid lines) depend weakly on the magnetic field even in the presence of significant TE-TM splitting $\kappa=1$ as in the calculation presented here. Note, that spin Meissner effect is not exact for states with non-zero net angular momentum $m_+ + m_-$ while it holds better for states with zero angular momentum such as $(0,0)$ and $(1,-1)$. Dashed lines mark constant amplitude solutions.
Nonlinearity parameters $\rho=3$ and $\alpha=-0.05$.
}
\end{center}
\end{figure}
\fi

\iffigures
\setlength{\unitlength}{0.1in}
\begin{figure*}
\begin{center}
$
\begin{array}{cc}
\begin{picture}(35,35)
\put(-1,0){\includegraphics[width=3.5in]{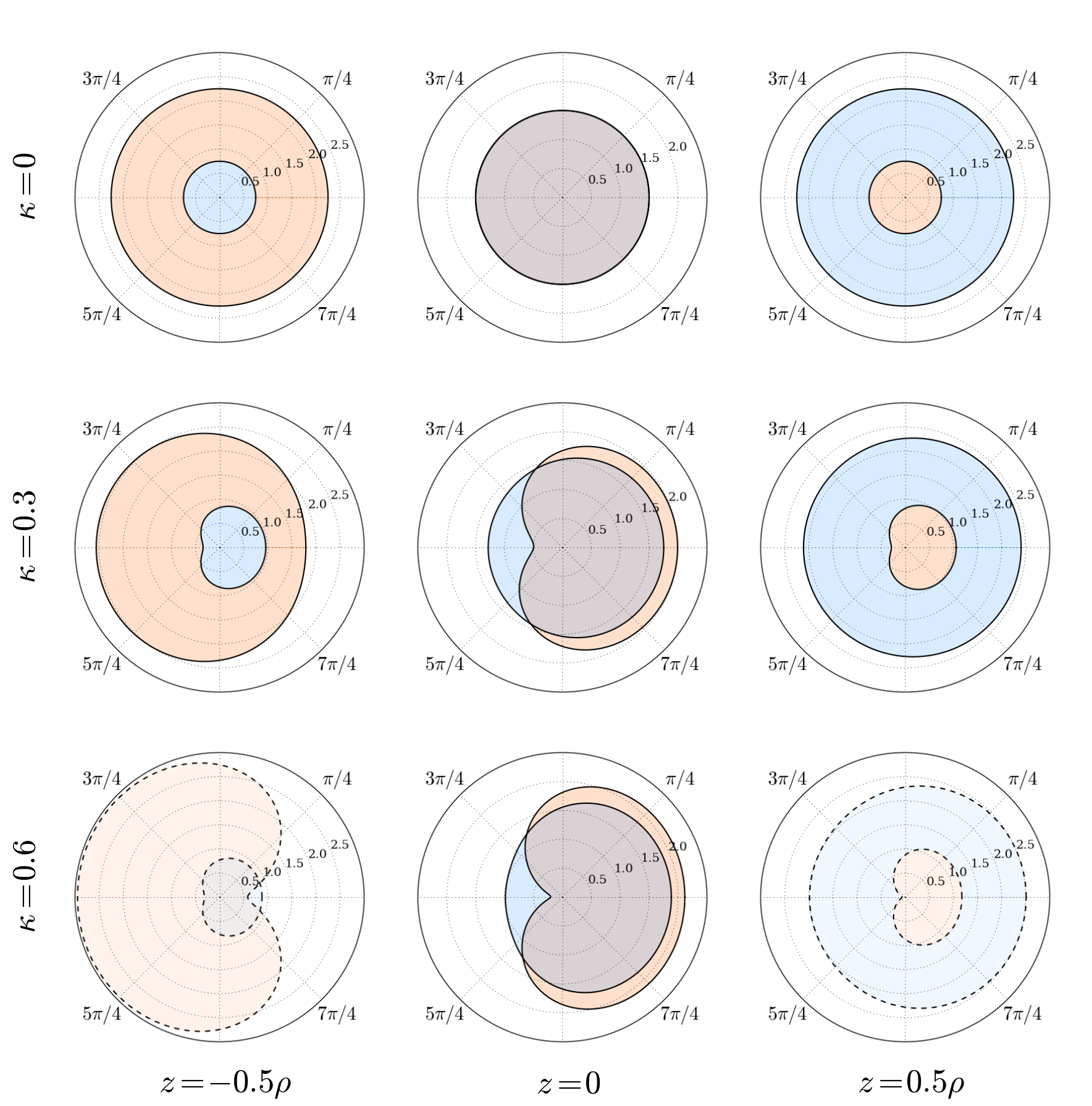}}
\put(0,0){(a)}
\end{picture}
 & 
\begin{picture}(35,35)
\put(0,0){\includegraphics[width=3.5in]{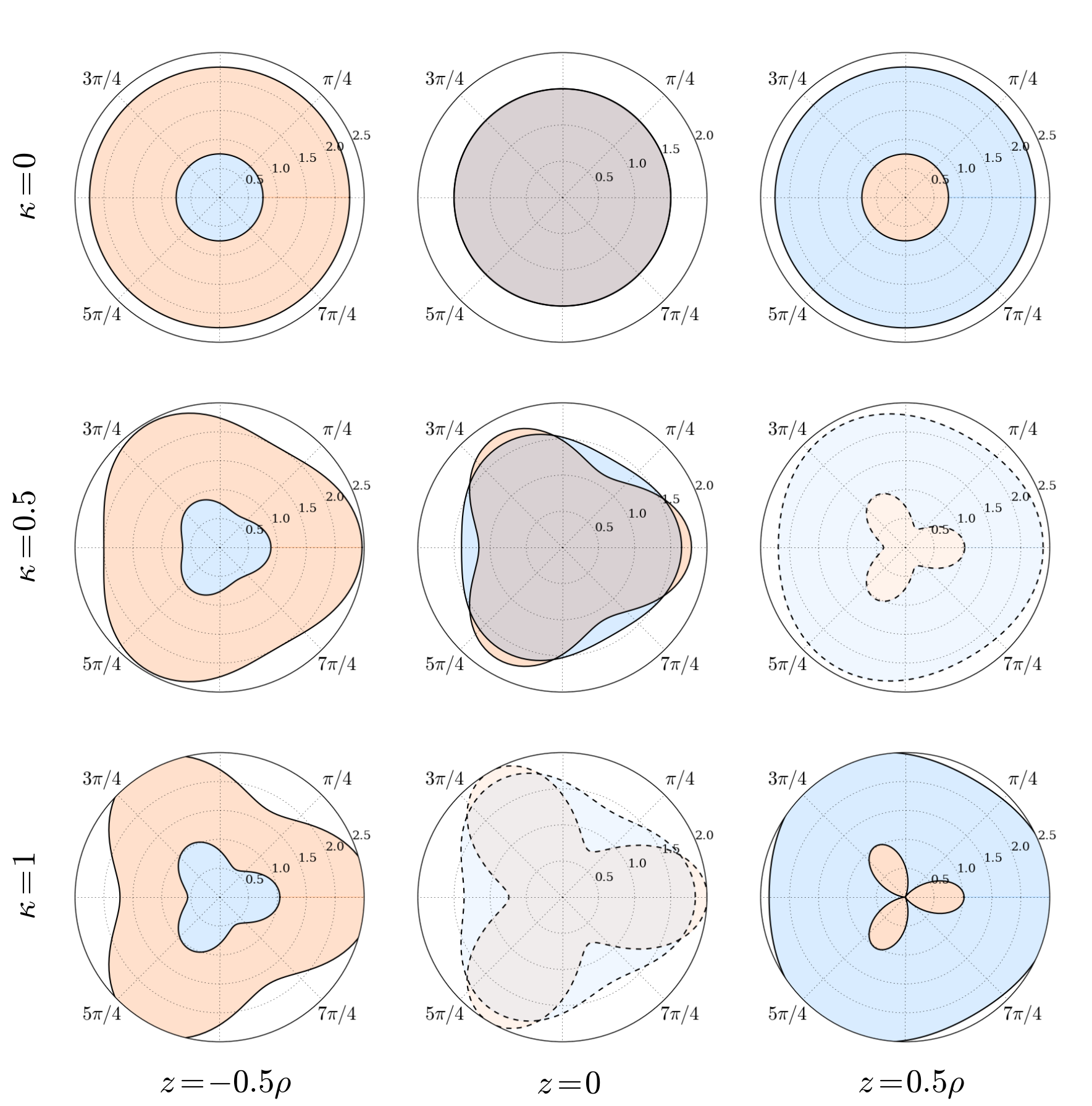}}
\put(1,0){(b)}
\end{picture}
\end{array}
$
\caption{\label{fig:hvs} 
(Color online) 
Topological spin Meissner effect in half-vortices $(-1,0)$ (a) and $(1,0)$ (b): the effect of magnetic field is balanced by changing densities of the spinor components of a vortex state, keeping the energy of the state nearly constant (see Fig.~\ref{fig:Omega-mu-rho3}).
Polar plots of numerically calculated density distributions are presented. Pink and blue colors represent densities $|\phi_+|^2$ and $|\phi_-|^2$, correspondingly. TE-TM splitting $\kappa$ takes values, from top to bottom: $0$, $0.3$ and $0.6$ in (a) and $0$, $0.5$ and $1.0$ in (b). Magnetic field is given by $z/\rho=-0.5$, $0$ and $0.5$ from left to right, with $z$ defined by eq.~\eqref{z}. Nonlinearity parameters $\rho=3$ and ${\alpha=-0.05}$. Dashed lines and fainter colors mark unstable states.
}
\end{center}
\end{figure*}
\fi

\iffigures
\setlength{\unitlength}{0.1in}
\begin{figure}
\begin{center}
\includegraphics[width=3.5in]{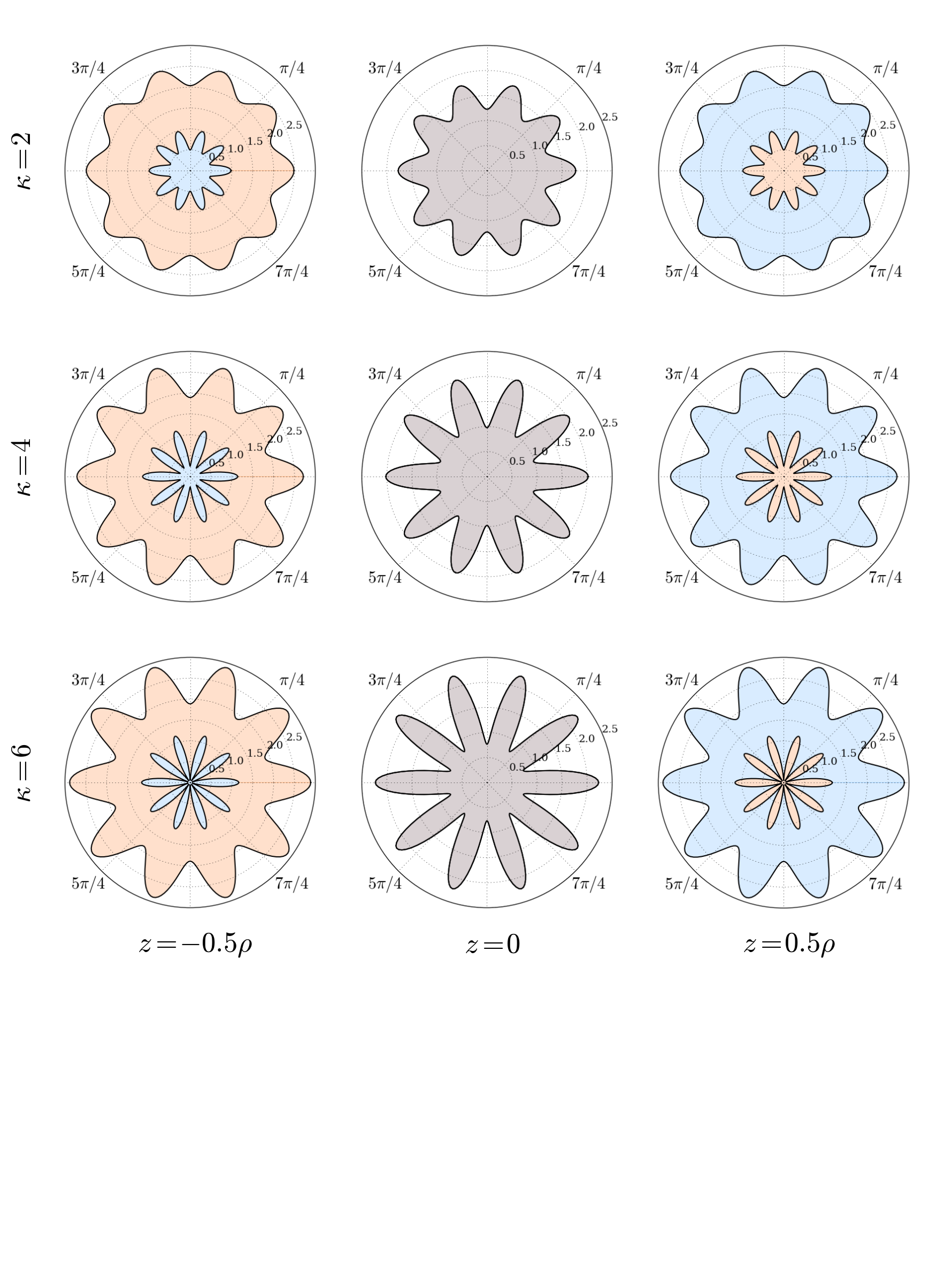}
\caption{\label{fig:star} 
(Color online) 
Numerically continued symmetry-breaking TSM state $(4,-4)$ of higher order rotational symmetry.  Pink and blue colors represent densities $|\phi_+|^2$ and $|\phi_-|^2$, correspondingly. The values of $\kappa$ are $2$, $4$ and $6$ from top to bottom. Magnetic field is given by $z/\rho=-0.5$, $0$ and $0.5$ from left to right, with $z$ defined by eq.~\eqref{z}. Balancing of the magnetic field by density distributions of the spinor components in a polarization vortex (topological spin Meissner effect) is seen here. This leads to all states having energies independent of the value of magnetic field. Nonlinearity parameters $\rho=3$ and ${\alpha=-0.05}$. All of the displayed configurations were found to be stable with respect to linear perturbations.
}
\end{center}
\end{figure}
\fi

\iffigures
\setlength{\unitlength}{0.1in}
\begin{figure*}
\begin{center}
$
\begin{array}{cc}
\begin{picture}(35,22)
\put(0,0){\includegraphics[width=3.5in]{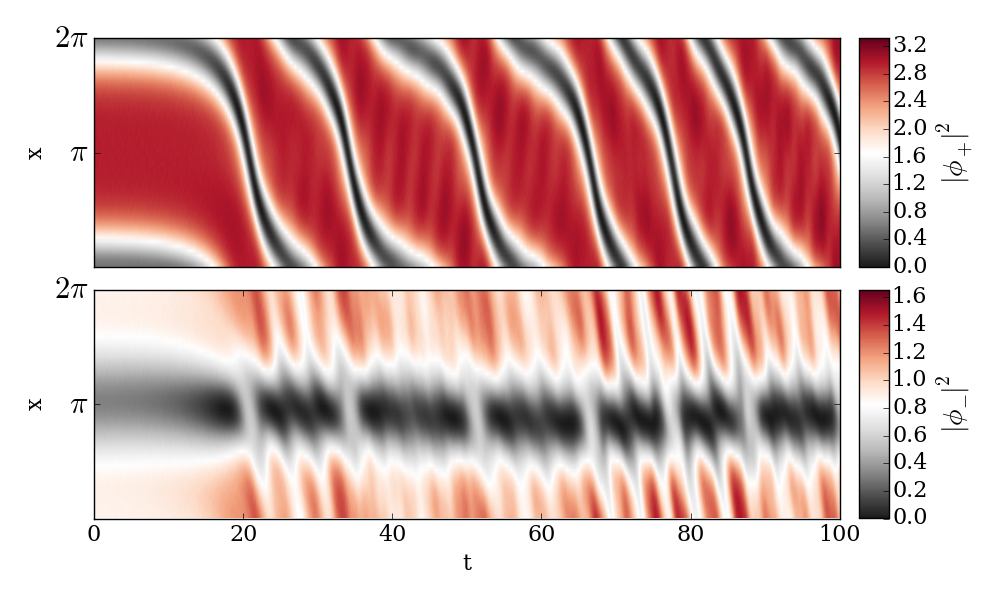}}
\put(3,0){(a)}
\end{picture}
&
\begin{picture}(35,22)
\put(0,0){\includegraphics[width=3.55in]{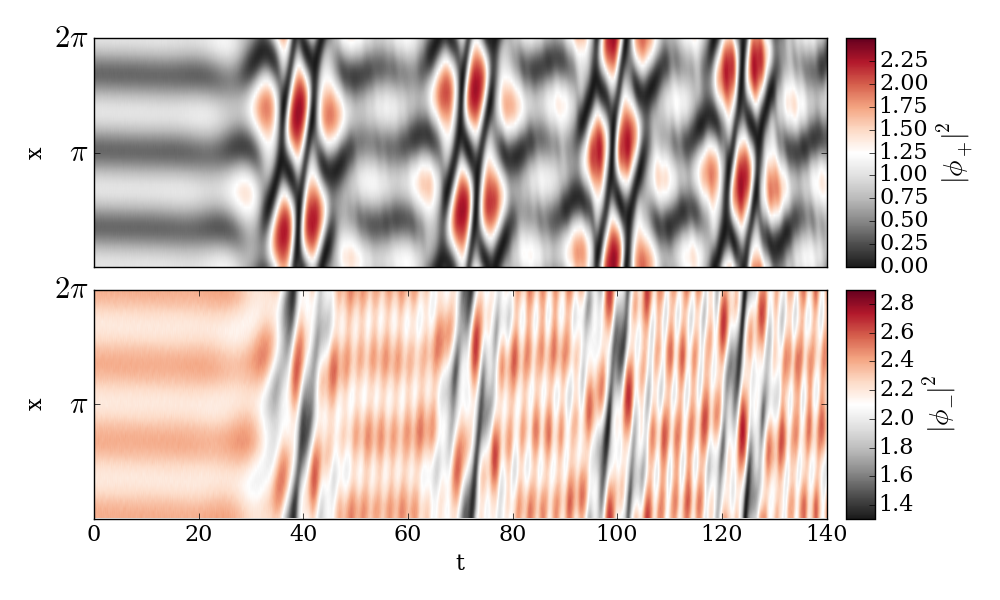}}
\put(3,0){(c)}
\end{picture}
\\
\begin{picture}(35,22)
\put(0,0){\includegraphics[width=3.5in]{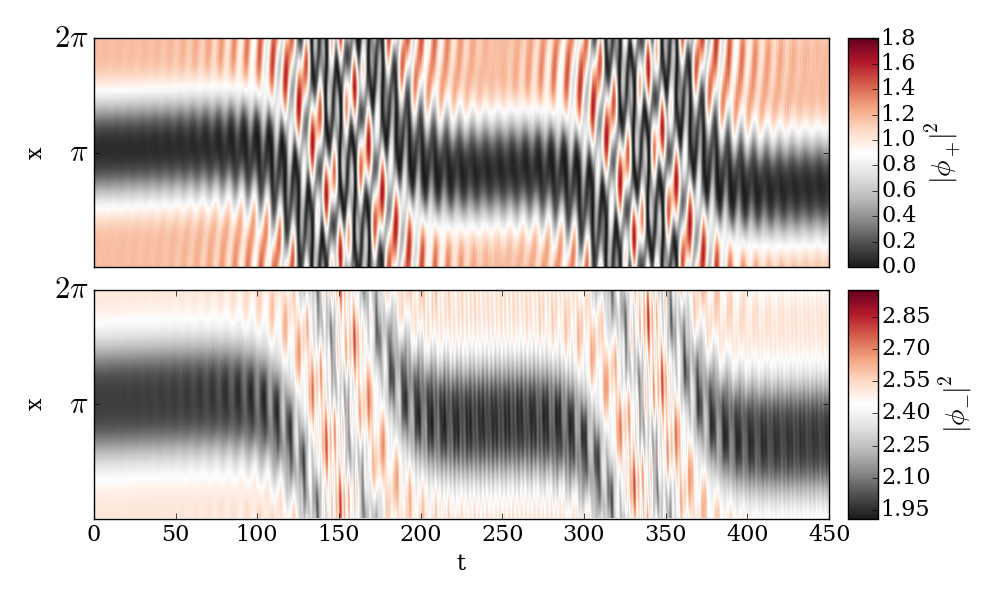}}
\put(3,0){(b)}
\end{picture}
&
\begin{picture}(35,22)
\put(0,0){\includegraphics[width=3.5in]{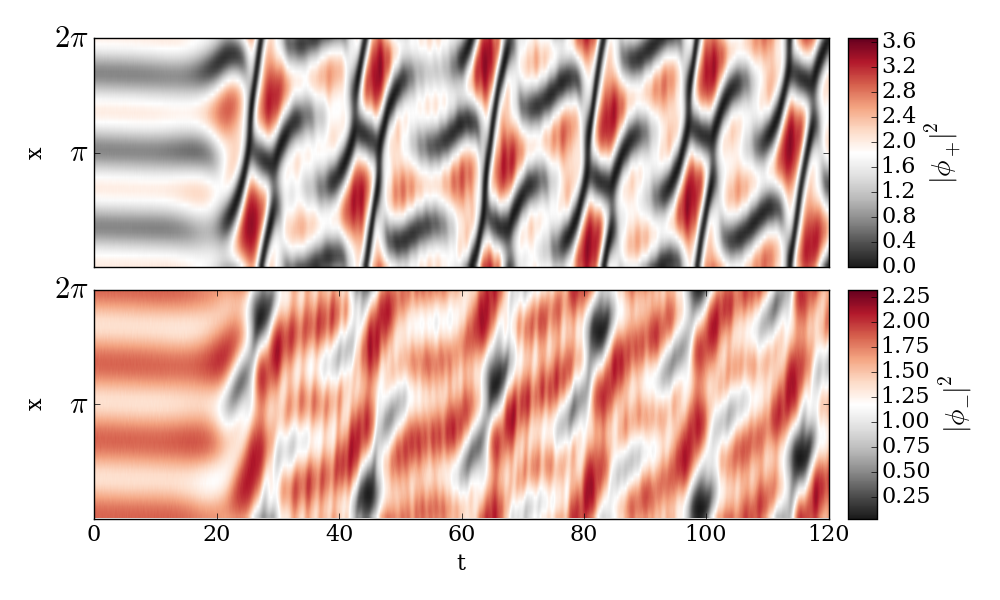}}
\put(3,0){(d)}
\end{picture}
\end{array}
$
\caption{\label{fig:instab-hvs} 
(Color online)
Dynamics of instabilities arising when an unstable half-vortices shown on Figures~\ref{fig:hvs} are distorted by a small initial perturbation: state $(-1,0)$ at (a) $z/\rho=-0.5$, $\kappa=0.6$  and (b) $z/\rho=0.5$, $\kappa=0.6$ (b) with density profile on Fig.~\ref{fig:hvs}a and state $(1,0)$ at (c) $z/\rho=0.5$, $\kappa=0.5$ and (d) $z/\rho=0$, $\kappa=1$ with density profile on Fig.~\ref{fig:hvs}b. Videos of animated dynamics are available in the Supplementary Material.
}
\end{center}
\end{figure*}
\fi

In the previous section we focused our attention to the TSM states with the constrained phase winding numbers $\Delta m = 2$: these are constant-amplitude solutions at non-zero TE-TM splitting. In this section we will investigate the fate of the more general states~\eqref{const-amp-sols} with arbitrary $m_+$, $m_-$. It turns out that solutions with $\Delta m \neq 2$ do not disappear in the presence of non-zero $\kappa$ but instead develop inhomogeneous profiles.

At $\kappa=0$ constant amplitude solutions of the auxiliary system of equations~\eqref{GP-phi} are expressed via solutions~\eqref{const-amp-sols} studied in Section~\ref{section:zero},
\begin{equation}
\phi_{\pm}(x)=\chi_{\pm}e^{i(m_{\pm}\pm 1) x}
\label{phi-pm}
\end{equation}

Important aspects of the continuation of TSM states from $\kappa=0$ to $\kappa\ne 0$ 
can be understood if we consider changes in the symmetry properties of the model equations and of the solutions themselves. Eqs.~\eqref{GP-phi} with $\kappa=0$ are invariant under the following three transformations: rotation of the total phase of the spinor, rotation of the relative phases of the spinor components and shift of the azimuthal coordinate $x$. 
However, not all of these operations are independent. Consider shift of the coordinate $x\to x + \Delta x$,
\begin{equation}
\mathcal{\hat R}\left(\Delta x\right)
\begin{pmatrix}
\chi_+ e^{i n_+ x}
\\
\chi_- e^{i n_- x}
\end{pmatrix}
= e^{i \Delta x (m_+ + m_-)/2 } 
\begin{pmatrix}
\chi_+ e^{i n_+ x - i \delta/2}
\\
\chi_- e^{i n_- x + i \delta/2}
\end{pmatrix}
\label{rotation}
\end{equation}
where $\delta=(\Delta m-2)\Delta x$ and $n_\pm = m_\pm \pm 1$. Thus the shift of the coordinate of the constant amplitude solutions~\eqref{phi-pm} is equivalent to rotation of the total phase, if $\Delta m = 2$, or, rotation of both total and relative phases, if $\Delta m \neq 2$.
When  TE-TM splitting is not zero, $\kappa\ne 0$, then the symmetry of the model with respect to the shift of the relative phase is broken. This affects very differently solutions with $\Delta m =2$ and $\Delta m\neq 2$. Those, with $\Delta m =2$ remain invariant with respect to rotations in $x$, which are equivalent to the corresponding shift in the still present total phase. Meanwhile former $\Delta m\ne 2$ solutions develop inhomogeneous density profiles as their rotational symmetry becomes broken.
Thus, for $\kappa\ne 0$, $\Delta m=2$ solutions have one broken symmetry (total phase) and one Goldstone mode associated with it and $\Delta m\ne 2$ solutions have two broken symmetries and two Goldstone modes.
While for $\kappa=0$, both types of solutions have two broken symmetries (in the total and relative phases). Because the number of Goldstone bosons does not change for $\Delta m\ne 2$ solutions, they do not branch as we introduce $\kappa\ne 0$, while $\Delta m=2$ split into branches, see Fig.~\ref{fig:kappa-mu-rho3}.

Numerical continuation of the solutions~\eqref{const-amp-sols} in parameter $\kappa$ to the domain of non-zero TE-TM splitting is presented on Fig.~\ref{fig:kappa-mu-rho3} for a fixed nonlinearity parameter $\rho=3$. At $\kappa=0$ energies of the solutions are given by Eqs.~\eqref{mu-plus},~\eqref{mu-minus} and~\eqref{mu-Meissner}, see also a cross section of Fig.~\ref{fig:families}b at $\rho=3$. 
With increasing TE-TM splitting parameter the solutions with winding numbers $(m_+,m_-)$, $\Delta m\neq 2$ continue to exist developing inhomogeneous density profiles. Splitting of the constant amplitude states with $\Delta m=2$ into two branches is also visible in Fig.~\ref{fig:kappa-mu-rho3}. Contrary to the constant amplitude states, symmetry breaking states do not split into branches.
Indeed, as can be seen from formula~\eqref{rotation}, states with different relative phase of the spinor components at $\kappa=0$ seed the same solution at $\kappa \neq 0$, apart from the coordinate shift and a common phase factor. Snapshots of evolution of densities of the spinor components under continuously changing $\kappa$ is shown on Fig.~\ref{fig:images-rho3}.

Symmetry-breaking states can be classified according to which states they can be continued from by increasing TE-TM splitting $\kappa$ from zero because they inherit topological properties of the seeding solutions, i.e. their two phase winding numbers. Indeed, the two topological invariants which could be used to characterize a symmetry-breaking TSM state are
\begin{equation}
\frac{1}{2\pi i} \int_0^{2\pi} \frac{\partial_x\phi_\pm}{\phi_\pm} dx
\end{equation}
coincide with $m_\pm$ of the constant amplitude TSM state at $\kappa=0$ to which it can be continued given that neither of the components turned to zero during the continuous transformation. 

To analyze analytically TSM states with broken rotational symmetry in presence of TE-TM splitting we use perturbative approach and consider a small distortion $\varepsilon_\pm(x)$ of the shape of the TSM state~\eqref{chi-Meissner},
\begin{equation}
\phi_\pm(x) = \left[\chi_\pm + \kappa \varepsilon_\pm(x) + O(\kappa^2)\right]e^{i (m_\pm \pm 1) x}
\end{equation}
\begin{equation}
\mu = \mu^{(0)} + \kappa \mu^{(1)} + O(\kappa^2)
\end{equation}
where $\varepsilon_{\pm}(x)$ are complex functions and $\chi_\pm$ are amplitudes of the TSM state at $\kappa=0$.
Substituting to~\eqref{GP-phi} we get the system of equations on $\varepsilon_{\pm}(x)$,
\begin{multline}
-\varepsilon_{\pm}'' - 2i m_{\pm} \varepsilon_{\pm}' 
 + \chi_{\pm}^2(\varepsilon_{\pm}+\varepsilon_{\pm}^*)+
\alpha\chi_{\pm}\chi_{\mp}(\varepsilon_{\mp}+\varepsilon_{\mp}^*)
\\
=\mu^{(1)} \chi_{\pm} - e^{\pm i (\Delta m-2) x} \chi_{\mp}
\label{epsilon}
\end{multline}
Assuming $\Delta m\neq 2$, we may seek for solutions of~\eqref{epsilon} in the form,
\begin{equation}
\varepsilon_{\pm} = A_{\pm}e^{\pm i(\Delta m-2)x} + B_{\pm}e^{\mp i(\Delta m-2)x} + C_{\pm}
\label{ABC}
\end{equation}
where coefficients $A_{\pm},B_{\pm}$ and $C_{\pm}$ can be chosen real. Substituting to~\eqref{epsilon} and using the normalization condition we get two decoupled systems of equations
\begin{equation}
\hat{\mathcal{H}}_{\Delta m-2}\mathbf{W} = \mathbf{R}
\label{PW}
\end{equation}
\begin{equation}
Q\mathbf{C} = 0
\label{QC}
\end{equation}
where $\mathbf{W}=(A_{+},B_{+},A_{-},B_{-})^T$, $\mathbf{C}=(C_{+},C_{-},\mu^{(1)})^T$, $\mathbf{R}=(-\chi_{-},0,-\chi_{+},0)^T$, matrix $\hat{\mathcal{H}}_{\Delta m-2}$ is given by Eq.~\eqref{H0} for $l=\Delta m-2$ and
\begin{equation}
Q=\begin{pmatrix}
2\chi_{+}^2 & 2\alpha\chi_+ \chi_- & -\chi_+ \\
2\alpha\chi_+\chi_- & 2\chi_{-}^2 & -\chi_- \\
\chi_+ & \chi_- & 0
\end{pmatrix}
\end{equation}
The determinant of matrix~$Q$ is $\det Q = 4\chi_+^2\chi_-^2(1-\alpha)$
which is non-zero as $\chi_-,\chi_+\neq 0$. Therefore, for the TSM states, only trivial solution to the system~\eqref{QC} exists, i.e. $C_1=C_2=\mu^{(1)}=0$. Because $\mu^{(1)}=0$ and l $\mu^{(0)}$ does not change with the magnetic field, energy $\mu$ of the symmetry-breaking solutions originated by TSM states is also independent of the magnetic field, i.e. {\it symmetry-breaking solutions originated by topological spin Meissner states remain spin Meissner states}, at least in the first order in TE-TM splitting.
Therefore, from~\eqref{mu-Meissner} for symmetry-breaking TSM states in presence of TE-TM splitting we have
\begin{equation}
\mu = \frac12 \left[ m_+^2+m_-^2 + \rho (1+\alpha) \right] + O(\kappa^2).
\end{equation}
To find the density profiles of the symmetry-breaking solutions we solve the system~\eqref{PW}. For small $\kappa$ perturbative solutions~\eqref{ABC} are in good agreement with our numerical calculations (see Supplementary Information for comparison between the theoretical and numerical results).

The influence of TE-TM splitting on the topological spin Meissner effect in shown on Figure~\ref{fig:Omega-mu-rho3}. As seen from the Figure, energies of symmetry-breaking TSM states depend weakly on the magnetic field even in the presence of significant TE-TM splitting $\kappa=1$. Notice that spin Meissner effect is not exact for states with non-zero net angular momentum $m_+ + m_-$ while it holds better for states with zero angular momentum such as $(0,0)$ and~$(1,-1)$.

Studies of topological spin Meissner effect in half-vortices $(-1,0)$ and $(1,0)$ are shown on Fig.~\ref{fig:hvs}a and b. As seen form the Figure, the magnetic field is balanced by the densities of the spinor components of a vortex state, with energy of the state remaining nearly constant (cf. Fig.~\ref{fig:Omega-mu-rho3}). 
Fig.~\ref{fig:Omega-mu-rho3}a and b shows polar plots of the numerically calculated densities $|\phi_+|^2$ and $|\phi_-|^2$ for a fixed nonlinearity parameter $\rho=3$. Dashed lines and fainter colors mark spectrally unstable states.
The other two half-vortices, $(0,1)$ and $(0,-1)$ coincide with $(1,0)$ and $(-1,0)$ when two circularly polarized are interchanged and direction of the magnetic field is reversed. 

As seen from the analytical formula~\eqref{ABC}, quantity $\Delta m - 2$ defines order of the discrete rotational symmetry in the density distribution of the states (i.e. number of peaks or deeps): in $(1,0)$ and $(0,-1)$ it has a three-fold rotational symmetry while in states $(0,1)$ and $(-1,0)$ the density distribution has "1-fold" symmetry (i.e. no rotational symmetry). 
Evolution of numerically calculated density distributions for a higher order state $(4,-4)$ with changing TE-TM splitting parameter $\kappa$ is shown on Fig.~\ref{fig:star}. Due to $|\Delta m - 2|=10$, its density distribution is 10-fold rotationally symmetric.

We investigate stability of the symmetry-breaking states numerically evaluating the eigenvalues of the discretized Hessian matrix (the continuous version of the Hessian matrix is given by Eq.~\eqref{Hessian} in Appendix~\ref{app:stab-th}). The unstable states resulting from this analysis are marked by dashed lines and faint colors in Figures~\ref{fig:hvs}a and~\ref{fig:hvs}b. No instabilities were found among the configurations displayed in Figure~\ref{fig:star}.
We analyzed the unstable states of half-vortices marked by dashed lines and faint colors in Fig.~\ref{fig:hvs}. The arising dynamics when an unstable state is disturbed by a small perturbation is shown on Fig.~\ref{fig:instab-hvs}. 
As seen from the Figure, in the initial stage the densities patterns are nearly constant as it takes time for the instability to develop. In the next stage when instability has grown large enough, quasi-periodic patterns appear which indicate an onset of propagating waves which modulate densities of the circular polarized components. Videos of the propagating density modulations is available as Supplementary Material.

\section{Discussion}

To conclude, we have shown that exciton-polariton condensate placed into a trap of non-simply connected geometry may exhibit states whose energies are independent of the applied magnetic field. Properties of these states are dictated by the topology of the condensate wavefunction, i.e. two phase winding numbers of its spinor components.
We analyzed the stability of these topological spin Meissner states and indicated the range of parameters where such states may exist and are stable. These findings helped us to shed the light onto the properties of half-vortices in a ring and gave a clue in understanding of the recent experiments.  We analyzed the effect of TE-TM splitting on the topological spin Meissner states and found that the stable states exist even in presence of significant TE-TM splitting strengths. Finally, we found that a certain class of TSM states exist which breaks rotational symmetry in presence of TE-TM splitting by developing inhomogeneous densities. 

The range of parameters discussed in this paper can be reached experimentally. Depending on the size of the ring and detuning, the characteristic energy $\hbar^2/2 m ^* R^2$ may be varied in a broad range of energies. For a ring diameter $10\;\mu m$  
the unit energy can be varies from $4\;\mu eV$ to $40\;\mu eV$, depending on the detuning. 
Therefore, both small $\Omega\sim 10$ and higher and well accessible in experiments. 
The effect of TE-TM splitting can be made significant, if desired.
In a $1~\mu m$ waveguide TE/TM splitting can be as high as $\sim1\;\rm meV$~\cite{Dasbach-PRB-2005,Kuther-PRB-1998} which allows to reach $\kappa\sim 10$ and even $\kappa\sim 100$, in normalized units. On the other hand $\kappa$ can be made negligibly small by choosing larger ring widths, controlling detuning~\cite{Duff-PRL-2015} and properties of distributed Bragg reflector~\cite{Panzarini-PRB-1999}.


\appendix
\section{TE-TM splitting in microcavity ring resonator}
\label{app:tetm}

The TE-TM splitting of linear polarization in quasi-one-dimensional microcavities may be of different physical origins such as difference in reflection coefficients for TE and TM polarizations in Bragg mirrors~\cite{Panzarini-PRB-1999}, anisotropy caused by thermal expansion~\cite{Dasbach-PRB-2005} and 
influence of the boundaries~\cite{Kuther-PRB-1998}.

The simpler case of $k$-independent TE-TM splitting may arise due to an anisotropy present in the system such as deformation due to a thermal stress~\cite{Dasbach-PRB-2005} or difference in boundary conditions for electric and magnetic fields at the cavity-to-air interface~\cite{Kuther-PRB-1998}. 
Assuming a homogeneously distributed asymmetry with the axis aligned along the radial/azimuthal direction of the ring, for the polaritons coupled to the TE and TM modes, the TE-TM Hamiltonian 
acting on the spinor wavefunction $(\psi_r(\varphi,t),\psi_\varphi(\varphi,t))^T$,
\begin{equation}
\hat{H}_{TE-TM}=
\begin{pmatrix}
-\Delta & 0
\\
0 & \Delta
\end{pmatrix}.
\label{HTETM-0}
\end{equation}
where $2\Delta$ is energy splitting for polarizations aligned along radial and azimuthal directions. Transforming to the basis of circular polarizations we get the Hamiltonian
\begin{equation}
\hat{H}_{TE-TM} = 
\begin{pmatrix}
0 & \Delta\, e^{-2i\varphi}
\\
\Delta\, e^{2i\varphi} & 0
\end{pmatrix}
\label{HTETM}
\end{equation}
acting on the wavefunction $(\psi_+(\varphi,t),\psi_-(\varphi,t))^T$. Here $\psi_+= (\psi_x \mp i \psi_y)/\sqrt{2}$ are components of a spinor  $\psi = \psi_+ \hat{e}_+ + \psi_- \hat{e}_-$ in the circular polarization basis with vectors $\hat{e}_\pm = (\hat{e}_x \pm i \hat{e}_y)/\sqrt{2}$. The TE-TM splitting Hamiltonian in the form~\eqref{HTETM} was established in work~\cite{Shelykh-PRL-2009} to describe polarization splitting of exciton-polariton condensate in one-dimensional ring interferometer. 

The $k$-dependent TE-TM splitting arises due to the property of distributed Bragg reflectors to have slightly different angular dispersions for TE and TM polarizations~\cite{Panzarini-PRB-1999}. This makes microcavity polaritons polarized longitudinally and transversely to the $\mathbf{k}$-vector 
to acquire different dispersion relations. 
In the effective mass approximation 
the $k$-dependent TE-TM splitting can be described by introducing two
masses $m_L$ and $m_T$ for polaritons polarized differently with respect to their propagation direction.
In a narrow ring resonator this type of TE-TM splitting becomes effectively independent of the wavenumber $k_\varphi$ along the ring, as long as $k_r\gg k_\varphi$ is satisfied. Indeed, consider exciton-polariton condensate confined in ring trap of radius $R$ and width $\Delta R$. With account of $k$-dependent TE-TM and Zeeman splittings, it can be described by a system if coupled spinor Gross-Pitaevskii equations~\cite{Maialle-PRB-1993,Flayac-PRB-2010,Solano-PRB-2014},
\begin{equation}
\begin{dcases}
\begin{split}
i\hbar\frac{\partial\Psi_{+}}{\partial t} = -\frac{\hbar^2}{2m^*}\nabla^2 \Psi_{+} 
+ \left( \alpha_1 |\Psi_{+}|^2 + \alpha_2 |\Psi_{-}|^2 \right)\Psi_{+} 
\\
+ \,\frac12 g_{eff} \mu_B B\,\Psi_{+}
- \beta \left(\partial_x - i \partial_y\right)^2 \Psi_{-}
\\
i\hbar\frac{\partial\Psi_{-}}{\partial t} = -\frac{\hbar^2}{2m^*}\nabla^2 \Psi_{-} 
+ \left( \alpha_1 |\Psi_{-}|^2 + \alpha_2 |\Psi_{+}|^2 \right)\Psi_{-} 
\\
- \,\frac12 g_{eff} \mu_B B\,\Psi_{-}
- \beta \left(\partial_x + i \partial_y\right)^2 \Psi_{+}
\end{split}
\end{dcases}
\label{GP-2D}
\end{equation}
where $\beta=\hbar^2(m_L^{-1}-m_T^{-1})/4$, $m^*=2m_T m_L/(m_T+m_L)$, 
$g_{eff}$ is the effective exciton-polariton $g$-factor, $\mu_B$ is the Bohr magneton and $B$ is the applied magnetic field,
$\alpha_1=U_0$ and $\alpha_2=U_0-2U_1$ are parameters characterizing polariton-polariton interactions. 
In the limit of a narrow ring we may use the adiabatic approximation and separate the radial dependence of the wavefunction, $\Psi_\pm(r,\varphi,t) = \zeta(r)\psi_\pm(\varphi,t)$ where $\zeta(r)$ is the normalized ground state wavefunction in the radial direction, satisfying the stationary Schr\"{o}dinger equation
$$
\frac{\hbar^2}{2m^*}\left(-\partial_r^2 - \frac{1}{r}\partial_r+\frac{1}{r^2}\right)
\zeta(r)=E\zeta(r)
$$
Neglecting lower order derivatives in $r$ 
we arrive to the following 1D model,
\begin{equation}
\begin{dcases}
\begin{split}
i\hbar\frac{\partial\psi_{+}}{\partial t} = -\frac{\hbar^2}{2m^*R^2}\frac{\partial^2}{\partial \varphi^2} \psi_{+} 
+ \left( \tilde\alpha_1 |\psi_{+}|^2 + \tilde\alpha_2 |\psi_{-}|^2 \right)\psi_{+} 
\\
+ \,\frac12 g_{eff} \mu_B B\,\psi_{+}
+ \tilde\beta e^{-2i\varphi} \psi_{-}
\\
i\hbar\frac{\partial\psi_{-}}{\partial t} = -\frac{\hbar^2}{2m^*R^2}\frac{\partial^2}{\partial \varphi^2} \psi_{-} 
+ \left( \tilde\alpha_1 |\psi_{-}|^2 + \tilde\alpha_2 |\psi_{+}|^2 \right)\psi_{-} 
\\
- \,\frac12 g_{eff} \mu_B B\,\psi_{-}
+ \tilde\beta e^{2i\varphi} \psi_{+}
\end{split}
\end{dcases}
\label{GP-1D}
\end{equation}
with effective parameters $\tilde\alpha_1$, $\tilde\alpha_2$ and $\tilde\beta$ which are connected to $\alpha_1$, $\alpha_2$ and $\beta$ via parameters of the ring. Note that the TE-TM splitting Hamiltonian obtained in Eq.~\eqref{GP-1D} is of the same form as given by the $k$-independent TE-TM splitting~\eqref{HTETM}. 

Introducing dimensionless units by rescaling the quantities entering Eqs.~\eqref{HTETM},~\eqref{GP-1D} to unit energy $E_0\equiv \hbar^2/2 m ^* R^2$, we obtain
\begin{equation}
\begin{dcases}
i\dot{\psi_+}= - \partial_\varphi^2\psi_+
+ \left( |\psi_+|^2 + \alpha |\psi_-|^2 \right)\psi_+ \\
\quad\quad\quad\quad\quad\quad + \Omega\psi_+  + \kappa e^{-2 i \varphi} \psi_-,
\\
i\dot{\psi_-} = - \partial_\varphi^2\psi_- + \left( |\psi_-|^2 + \alpha |\psi_+|^2  \right)\psi_- \\
\quad\quad\quad\quad\quad\quad - \Omega\psi_- + \kappa e^{2 i \varphi} \psi_+.
\end{dcases}
\end{equation}
where $\alpha\equiv\tilde\alpha_2/\tilde\alpha_1=\alpha_2/\alpha_1$, $\Omega\equiv \frac12 g_{eff} \mu_B B/E_0$, $\kappa\equiv \tilde\beta/E_0$ or
$\kappa\equiv \Delta/E_0$ depending on the origin of TE-TM splitting.
The number of particles is given by $N=2\pi\rho E_0/\tilde\alpha_1$ where $\rho\equiv \frac{1}{2\pi}\int_0^{2\pi}\left(|\psi_+|^2+|\psi_-|^2\right)d\varphi$. We use $x$ for $\varphi$ in the main text to simplify the notation.

The early experimental and theoretical attempts~\cite{Renucci,Kasprzak,Vladimirova,Voros,Tassone,Combescot,Schumacher,Glazov,Wouters} to estimate $\alpha_1$ and $\alpha_2$
generally agree that $\alpha_1>0$ and $\alpha_1\gg|\alpha_2|$ with some works suggesting negative $\alpha_2$. The recent investigations~\cite{Vladimirova-PRB-2010, Takemura-PRB-2014} of the ratio $\alpha_2/\alpha_1$ have shown that the ratio depends significantly on the detuning~$\delta$ between exciton and photon modes and may change from very negative (smaller than $-1$ for small negative~$\delta$) to positive values (for larger~$\delta$). In our calculations throughout this paper we use a ``conservative" estimate $\alpha=-0.05$.

\section{Stability analysis: numerical calculations}
\label{app:stab-num}

We analyze stability of the constant-amplitude solutions against the Bogoliubov-de Gennes excitations, see, e.g.~\cite{Skryabin-PRA-2000}. 
Consider a small time-dependent perturbation $\varepsilon_{\pm}(x,t)$ around a stationary constant amplitude solution,
\begin{equation}
\phi_{\pm}(x,t)=\left[\chi_{\pm}+\varepsilon_{\pm}(x,t)\right]e^{i n x}.
\end{equation}
Substituting to~\eqref{GP-phi} we get a system of linear equations on $\varepsilon_{\pm}(x,t)$,
\begin{multline}
i\dot\varepsilon = -\varepsilon_{\pm}'' - 2i(n\mp 1) \varepsilon_{\pm}' 
 + \chi_{\pm}^2(\varepsilon_{\pm}+\varepsilon_{\pm}^*)+
\alpha\chi_{\pm}\chi_{\mp}(\varepsilon_{\mp}+\varepsilon_{\mp}^*)
\\
+ \left[-\mu+n^2 + \chi_{\pm}^2+\alpha\chi_{\mp}^2 \pm(\Omega-2n)\right] \varepsilon_{\pm}+\kappa\varepsilon_{\mp} =0
\end{multline}
Expanding $\varepsilon_{\pm}(x,t)$ into the Fourier series in $x$,
\begin{equation}
\varepsilon_{\pm}(x,t) = \sum_{l=-\infty}^{\infty} U_{\pm,l}(t) e^{ilx} + V_{\pm,l}^*(t) e^{-ilx}
\end{equation}
we get a series of decoupled systems of equations parametrized by an integer $l$.
To analyze the stability we solve the eigenvalue problem
\begin{equation}
\hat\eta \mathcal{H}\mathbf{W}=\lambda \mathbf{W}
\end{equation}
with $\mathbf{W}=(U_{+,l},V_{+,l},U_{-,l},V_{-,l})$,
\begin{equation}
\hat\eta=
\begin{pmatrix}
1 & 0 & 0 & 0 \\
0 & -1 & 0 & 0 \\
0 & 0 & 1 & 0 \\
0 & 0 & 0 & -1
\end{pmatrix}
\end{equation}
and
\begin{equation}
\mathcal{H}=\begin{pmatrix}
d_+ & \chi_+^2 & \alpha\chi_+\chi_- +\kappa & \alpha\chi_+\chi_-
\\
\chi_+^2 & \tilde{d}_+ & \alpha\chi_+\chi_- & \alpha\chi_+\chi_- +\kappa
\\
\alpha\chi_+\chi_- +\kappa & \alpha\chi_+\chi_- & d_- & \chi_-^2
\\
\alpha\chi_+\chi_- & \alpha\chi_+\chi_- +\kappa & \chi_-^2 & \tilde{d}_-
\end{pmatrix}
\end{equation}
where $A_{\pm}$ and $B_{\pm}$ are given by
\begin{multline}
d_\pm \equiv 
l^2 + 2l (n\mp 1) + \chi_{\pm}^2 \\
+ \left[-\mu+m^2 + \chi_{\pm}^2+\alpha\chi_{\mp}^2 \pm(\Omega-2n)\right]
\end{multline}
\begin{multline}
\tilde{d}_\pm \equiv 
l^2 - 2l (n\mp 1) + \chi_{\pm}^2 \\
+ \left[-\mu+m^2 + \chi_{\pm}^2+\alpha\chi_{\mp}^2 \pm(\Omega-2n)\right].
\end{multline}
The solution is spectrally unstable if there is at least one eigenvalue with positive imaginary part $\Im\lambda > 0$.

\section{Stability analysis: theory}
\label{app:stab-th}

Consider a small time-dependent perturbation $\ep_{\pm}(x,t)$ around a stationary (in general, $x$-dependent) solution $\phi_{\pm}(x)$.
For $\pmb\ep=(\ep_+,\ep_+^*,\ep_-,\ep_-^*)^T$ we get the equation
\begin{equation}
i\dot{\pmb\ep}=\hat L\pmb \varepsilon
\label{eps-dynamic}
\end{equation}
where $\hat L=\hat\eta\hat H$, where $\hat\eta$ was defined above and
\begin{widetext}
\begin{equation}
\hat H=\left(\begin{array}{cccc} 
\hat D_+ +2|\phi_+|^2 +\al |\phi_-|^2 & \phi_+\phi_+ &\al \phi_-^*\phi_++\kappa &\al \phi_-\phi_+  
\\ \phi_+^*\phi_+^*&\hat D_+^* +2|\phi_+|^2 +\al |\phi_-|^2  &\al \phi_-^*\phi_+^* & \al \phi_-\phi_+^*+\kappa 
\\ \al \phi_+^*\phi_-+\kappa &\al \phi_+\phi_-& \hat D_- +2|\phi_-|^2 +\al |\phi_+|^2 & \phi_-\phi_- 
\\ \al \phi_+^*\phi_-^* & \al \phi_+\phi_-^*+\kappa &  \phi_-^*\phi_-^*&\hat D_-^* +2|\phi_-|^2 +\al |\phi_+|^2  \end{array}\right)
\label{Hessian}
\end{equation}
\end{widetext}
Substituting $\pmb \ep (x,t) = \pmb\ep(x) e^{-i\la t}$ to~\eqref{eps-dynamic} we get
\begin{equation}
\hat L\pmb\ep=\la\pmb\ep
\end{equation}
Operator $\hat L$ has the following properties:
\begin{equation}
\hat L\pmb q_1=0,~\hat L\pmb q_2=\pmb q_1,~\hat L\hat L\pmb q_2=0
\end{equation}
where 
\begin{equation}
\pmb q_1=\left(\begin{array}{c} \phi_+\\-\phi_+^*\\ \phi_-\\-\phi_-^*\end{array}\right),~\pmb q_2={\p\over\p\mu}\left(\begin{array}{c} \phi_+\\ \phi_+^*\\ \phi_-\\ \phi_-^*\end{array}\right)
\end{equation}
are two zero modes of operator $\hat L^2$. Define the operator $\hat L^{\dagger}=\hat H\hat\eta$ conjugated to $L$ with respect to the dot product
\begin{equation}
\braket{\pmb f,\pmb g} \equiv \int_0^{2\pi} \pmb f^*(x) \cdot \pmb g(x) \; dx
\end{equation}
For $\hat L^\dagger$ we have
\begin{equation}
\hat L^{\dagger}\pmb Q_1=0,~\hat L^{\dagger}\pmb Q_2=\pmb Q_1,~\hat L^{\dagger}\hat L^{\dagger}\pmb Q_2=0
\end{equation}
where $\pmb Q_{1,2}=\hat\eta\,\pmb q_{1,2}$.

Suppose $\phi_{\pm}$ are are calculated at $\ka=0$. Then there exists $p_1$ as well,
\begin{equation}
\hat L_0\pmb p_1=0,~\hat L_0\pmb p_2=\pmb p_1,~\hat L_0\hat L_0\pmb p_2=0
\end{equation}
where $\pmb p_1$ and $\pmb p_2$ are given by
\begin{equation}
\pmb p_1=\left(\begin{array}{c} -\phi_+\\ \phi_+^*\\ \phi_-\\ -\phi_-^*\end{array}\right),~
\pmb p_2={\p\over\p\Omega}\left(\begin{array}{c} \phi_+\\ \phi_+^*\\ \phi_-\\ \phi_-^*\end{array}\right)
\end{equation}
and
\begin{equation}
\hat L_0^{\dagger}\pmb P_1=0,~\hat L_0^{\dagger}\pmb P_2=\pmb P_1,~\hat L_0^{\dagger}\hat L_0^{\dagger}\pmb P_2=0
\end{equation}
with $\pmb P_{1,2}=\hat\eta\,\pmb p_{1,2}$.

In the case when the state is split into two branches at $\kappa\neq 0$, we use perturbative expansion for $\la$ a square root of the series in $\kappa$ for eigenvalue of operator $\hat L^2$,
\begin{equation}
\hat L=\hat L_0+\kappa\hat L_1+O(\ka^2)
\end{equation}
\begin{equation}
\la=\ka^{1/2}\left[ \la_0+\la_1\kappa +O(\kappa^2)\right]
\end{equation}
\begin{equation}
\pmb\ep=\pmb\ep_0+\ka^{1/2}\pmb\ep_1+\kappa\pmb\ep_2+O(\kappa^{3/2})
\end{equation}
In the first orders we have
\begin{equation}\hat L_0\pmb\ep_0=0\end{equation}
\begin{equation}\hat L_0\pmb\ep_1=\la_0\pmb \ep_0\end{equation}
\begin{equation}\hat L_0\pmb\ep_2+ \hat L_1\pmb\ep_0=\la_0\pmb \ep_1\end{equation}
Applying $L_0$ from the left to the last equation,
\begin{equation}
\hat L_0^2 \pmb\varepsilon_2 + \hat L_0 \hat L_1 \pmb\varepsilon_0=\lambda_0^2\pmb\varepsilon_0
\end{equation}
and forming a dot product with $\pmb P_2$, $\pmb Q_2$,
\begin{equation}
\begin{cases}
\braket{\pmb P_2,\hat L_0^2\pmb \varepsilon_2} + \braket{\pmb P_2,\hat L_0 \hat L_1 \pmb \varepsilon_0}=\lambda_0^2\braket{P_2,\pmb \varepsilon_0}\\
\braket{\pmb Q_2,\hat L_0^2\pmb \varepsilon_2} + \braket{\pmb Q_2,\hat L_0 \hat L_1 \pmb \varepsilon_0}=\lambda_0^2\braket{\pmb Q_2,\pmb \varepsilon_0}
\end{cases}
\end{equation}
\begin{equation}
\begin{cases}
\braket{\pmb P_1,\hat L_1 \pmb \varepsilon_0}=\lambda_0^2\braket{\pmb P_2,\pmb \varepsilon_0}\\
\braket{\pmb Q_1,\hat L_1 \pmb \varepsilon_0}=\lambda_0^2\braket{\pmb Q_2,\pmb \varepsilon_0}
\end{cases}
\end{equation}
Due to the degeneracy we take the linear superposition 
\begin{equation}
\pmb\varepsilon_0 = a\pmb p_1 +b\pmb q_1
\label{eps0}
\end{equation}
with constant coefficients $a$ and $b$,
Thus,
\begin{equation}
\begin{cases}
a\braket{\pmb P_1,\hat L_1 \pmb p_1}+b\braket{\pmb P_1,\hat L_1 \pmb q_1}
-\lambda_0^2 \left[ a\braket{P_2|p_1} + b\braket{P_2,q_1}\right]=0\\
a\braket{\pmb Q_1,\hat L_1 \pmb p_1}+b\braket{\pmb Q_1,\hat L_1 \pmb q_1}
-\lambda_0^2 \left[ a\braket{\pmb Q_2|\pmb p_1} + b\braket{\pmb Q_2,\pmb q_1}\right]=0
\end{cases}
\end{equation}
Also, as $\hat L(\kappa) \pmb q_1(\kappa)=0$ for arbitrary $\kappa$, there exist $\tilde{\pmb q_1}$ 
and 
$\tilde{\pmb Q_1}$ 
such that $\hat L_1 \pmb{q_1} = \hat L_0 \pmb{\tilde{q}_1}$ and $\hat L_1^\dagger \pmb{Q_1} = \hat L_0^\dagger \pmb{\tilde{Q}_1}$. 
Therefore, $\braket{\pmb P_1,\hat L_1 \pmb q_1}=\braket{\pmb Q_1,\hat L_0 \tilde{\pmb q}_1}=0$, $\braket{\pmb Q_1,\hat L_1 \pmb q_1}=\braket{\pmb Q_1,\hat L_0 \tilde{\pmb q}_1}=0$ and 
$\braket{\pmb Q_1,\hat L_1 \pmb p_1}=\braket{\tilde{\pmb Q}_1,\hat L_0 \pmb p_1}=0$.
\begin{equation}
\begin{cases}
a\left(\braket{\pmb P_1,\hat L_1 \pmb p_1} - \lambda_0^2 \braket{\pmb P_2,\pmb p_1}\right) 
- \lambda_0^2 b\braket{\pmb P_2,\pmb q_1}=0\\
a\lambda_0^2 \braket{\pmb Q_2,\pmb p_1} 
+ \lambda_0^2 b\braket{\pmb Q_2,\pmb q_1}=0
\end{cases}
\label{ab-system}
\end{equation}
Thus, for the two non-zero eigenvalues $\lambda_0$ we have the equation
\begin{equation}
\lambda_0^2 = \frac{\braket{\pmb P_1,\hat L_1 \pmb p_1}\braket{\pmb Q_2,\pmb q_1} }{\braket{\pmb Q_2,\pmb q_1}\braket{\pmb P_2,\pmb p_1} - \braket{\pmb Q_2,\pmb p_1}\braket{\pmb P_2,\pmb q_1}}
\end{equation}
Evaluating $\pmb p_1$,$\pmb q_1$, $\pmb P_2$,$\pmb Q_2$ for the TSM state $\Delta m =2$ at $\kappa=0$ (see Eqs.\eqref{chi-Meissner}, \eqref{condition-Meissner} and~\eqref{mu-Meissner}) we get: $\braket{\pmb Q_2,\pmb p_1}=0$, $\braket{\pmb P_2,\pmb q_1}=0$,  $\braket{\pmb P_2,\pmb p_1}=4\pi/(1-\alpha)$, $\braket{\pmb Q_2,\pmb q_1}=4\pi/(1+\alpha)$. 
Therefore, for $\lambda_0^2$ we have
\begin{equation}
\lambda_0^2 = \frac{\braket{\pmb P_1,\hat L_1 \pmb  p_1} }{\braket{\pmb P_2,\pmb p_1} } = \frac{(1-\alpha)}{4\pi}\braket{\pmb p_1,\hat H_1 \pmb p_1}
\label{l02}
\end{equation}
Operator $H_1$ can be found by substituting perturbative solution for TSM state to~\eqref{Hessian}. Evaluating~\eqref{l02} we get the formula~\eqref{lambda02}.

From the system~\eqref{ab-system} we find coefficients $a$ and $b$ which define the instability mode~\eqref{eps0}. For the considered case we get $b=0$ and therefore the unstable mode is 
$\varepsilon_{\pm} \sim \pm \chi_\pm e^{inx}$.

\begin{acknowledgments}
This work is supported under the project RFMEFI58715X0020 of the Federal Targeted Programme Research and Development in Priority Areas of Development of the Russian Scientific and Technological Complex for 2014-2020 of the Ministry of Education and Science of Russia. We acknowledge support from the
European Commission FP7 LIMACONA and H2020 SOLIRING projects.
I.A.S. acknowledges support from the Singaporean Ministry of Education under AcRF Tier 2 grant MOE2015-T2-1-055 and from Rannis Excellence Project ``2D transport in the regime of strong light-matter coupling".  D.V.S. acknowledges support from the Leverhulme trust.
\end{acknowledgments}

\end{document}